\PassOptionsToPackage{pdfpagelabels=false}{hyperref} 
\documentclass[a4paper,fleqn,usenatbib]{mnras}

\usepackage[T1]{fontenc}
\usepackage{ae,aecompl}

\usepackage{graphicx}	
\usepackage{amssymb}	
\usepackage{amsmath}	
\usepackage{siunitx}
\usepackage{grffile} 
\usepackage{subcaption}
\usepackage{multicol,multirow,booktabs}
\usepackage{hyperref}
\newcommand{\xismu}{$\xi(s,\mu)\,$} 
\newcommand{\xiproj}{$\xi_{\perp}(s)\,$}
\newcommand{\xisigpi}{$\xi(\sigma,\pi)\,$}
\newcommand{\sigmaone}{$\sigma_{0}=0.01\,$} 
\newcommand{\sigmatwo}{$\sigma_{0}=0.02\,$} 
\newcommand{\sigmathree}{$\sigma_{0}=0.03\,$} 

\def\be{\begin{equation}}
\def\ee{\end{equation}}
\def\ba{\begin{eqnarray}}
\def\ea{\end{eqnarray}}

\def\nn{\nonumber}

\title[Calculating the correlation function using the wedge approach]{Cosmic distance 
	   determination from photometric redshift samples using BAO peaks only}

\author[Srivatsan Sridhar, Yong-Seon Song]{
Srivatsan Sridhar$^{1}$\thanks{E-mail: srivatsan@kasi.re.kr}
and
Yong-Seon Song$^{1}$
\\
$^{1}$Korea Astronomy \& Space Science Institute 776, Daedeokdae-ro, Yuseong-gu, Daejeon,
Republic of Korea (34055)\\
}

\date{}

\pubyear{2018}

\begin{document}
\label{firstpage}
\pagerange{\pageref{firstpage}--\pageref{lastpage}}
\maketitle

\begin{abstract}
The galaxy distributions along the line-of-sight are significantly contaminated by the 
uncertainty on redshift measurements obtained through multiband photometry, which makes it 
difficult to get cosmic distance information measured from baryon acoustic oscillations, 
or growth functions probed by redshift distortions. 
We investigate the propagation of the uncertainties into large scale clustering
by exploiting all known estimators, and propose the wedge approach as a 
promising analysis tool to extract cosmic distance information still remaining in the 
photometric galaxy samples. 
We test our method using simulated galaxy maps with photometric uncertainties of 
$\sigma_{0} =\left(0.01, 0.02, 0.03\right)$. 
The measured anisotropy correlation function $\xi$ is binned into the radial direction of $s$ 
and the angular direction of $\mu$, and the variations of \xismu with perpendicular and radial 
cosmic distance measures of $D_A$ and $H^{-1}$ are theoretically estimated by an improved RSD 
model. Although the radial cosmic distance $H^{-1}$ is unable to be probed from any of the 
three photometric galaxy samples, the perpendicular component of $D_A$ is verified to be 
accurately measured even after the full marginalisation of $H^{-1}$. We measure $D_A$ with 
approximately 6\% precision which is nearly equivalent to what we can expect from spectroscopic
DR12 CMASS galaxy samples.
\end{abstract}




\section{Introduction}

Since the discovery of the cosmic acceleration \citep{Riess_1998,Perlmutter_1999}, 
many theoretical models have been proposed to explain the cause of it by introducing a 
positive cosmological constant, a time varying dark energy component, or a modified theory 
of gravity. As most ongoing observations support the $\Lambda$CDM model with the presence of 
the cosmological constant, it becomes an interesting observational mission to confirm 
$\Lambda$CDM in high precision, or to probe any possible deviation from it. The expansion 
history of the Universe can be revealed by diverse cosmic distance measures in tomographic 
redshift space, such as cosmic parallax \citep{Parallax}, standard 
candles \citep{Standard_candle} or standard rulers \citep{Eisenstein_1998,Eisenstein_2005}, 
and the possible presence of dynamical dark energy evolution can be confirmed or excluded 
in precision.

The tension between gravitational infall and radiative pressure caused by the baryon-photon 
fluid in the early Universe gave rise to an acoustic peak structure which was imprinted on the 
last-scattering surface (hereafter BAO) \citep{Peebles_Yu}. BAO is known as a relatively risk 
free standard ruler technique to probe cosmic distances. The BAO feature has been measured 
through the correlation function \citep{Blake_2003,Eisenstein_2005}, and the most successful 
measurements in the clustering of large-scale structure at low redshifts have been obtained 
using data from SDSS \citep{Eisenstein_2005,Estrada_2009,Padmanabhan_2012,Hong_2012,Veropalumbo_2014,Veropalumbo_2016,Alam_2017}.
In the near future, the wider and deeper Dark Energy Spectroscopic Instrument (hereafter DESI) 
survey will be launched to 
probe the earlier expansion history with greater precision using spectroscopic redshifts.
However, the footprint photometric 
survey for DESI has already been completed. Although these photometric redshifts are measured 
with a much poorer resolution, there might still be possible BAO signatures that have not been
contaminated by the redshift uncertainty. If that is the case, we should be able to provide the
precursor of cosmic distance information which will be revealed by the follow up spectroscopy 
experiment much later on. We investigate the optimised methodology to extract the 
uncontaminated cosmic distance information in the photometric data sets. This statistical tool
can also be applicable for many imaging surveys, such as the the ongoing 
Dark Energy Survey \citep{DES}, the PAUS survey \citep{Benitez_2009_PAU,PAU}, the 
Javalambre Physics of the Accelarating Universe Astrophysical survey
(JPAS) \citep{JPAS}
or upcoming surveys such as Large Synoptic Survey Telescope \citep{LSST_1,LSST_2} and 
\textit{Euclid} \citep{euclid_red_book}.

\begin{figure*}
	\includegraphics[width=\textwidth]{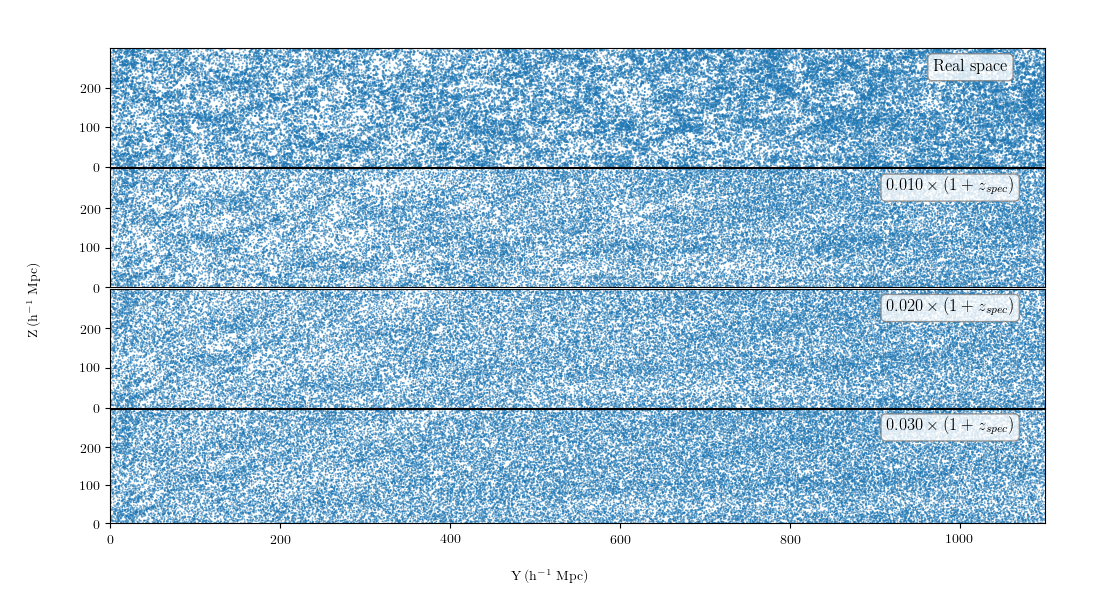}   
	\caption{The galaxy distributions with varying uncertainties of $z$ measurement 
			 are presented along the transverse and radial directions. The maps with 
			 different $z$ dispersion of $\sigma_{0}=(0,0.01,0.02,0.03)$ are shown from 
			 the top to the bottom panels. Here $Y$ and $Z$ denote the transverse and 
			 radial coordinates respectively.} 
	\label{fig:Y_vs_Z_plot}   
\end{figure*}

It is known that the correlation along the line-of-sight (hereafter LOS) is obtained with 
precise spectroscopic redshift measurements with the dispersion being in the order of 
$\sigma_{z}/(1+z) < 0.001$. This dispersion has decreased with state-of-the-art 
spectrograph design with the upcoming DESI survey 
\citep{DESI_1} which is expected to have $\sigma_{z}/(1+z) \sim 0.0005$, but such surveys 
are time consuming. If the imaging survey is done with multiple bands, then the photometric 
redshifts can be estimated as precise as $\sigma_{z}/(1+z) > 0.01$. 
It has been recently shown by the Physics of the Accelerating Universe Survey (PAUS)
that by using a filter system with 40 filters, each with a width close to 100$\si{\angstrom}$, 
the redshifts dispersion 
can be further reduced to $\sigma_{z}/(1+z) \sim 0.003$ \citep{Benitez_2009_PAU,Marti_PAU}. 
The ongoing Dark Energy 
Survey aims to cover about 5000 deg$^{2}$ of the sky  with a photometric accuracy of 
$\sigma_{z} \approx 0.08$ out to $z\approx1$ \citep{Sanchez_DES}. 
Future surveys such as LSST 
and \textit{Euclid} are expected to make a significant leap forward. The \textit{Euclid} Wide 
Survey, planned to cover 15000 deg$^{2}$, is expected to deliver photometric redshifts with 
uncertainties lower than $\sigma_{z}/(1+z) < 0.05$, and possibly  $\sigma_{z}/(1+z) < 0.03$, 
over the  redshift range [0,2] \citep{euclid_red_book}. While the photo-z survey provides 
more observed galaxies compared to a spectroscopic survey even at deeper redshifts, an 
unpredictable damping of clustering at small scales and a smearing of the BAO peak is caused 
by the photo-z uncertainty~\citep{Estrada_2009}.
Thus the cosmological information contained in the large-scale clustering is expected to 
be significantly contaminated. However, the cosmic distance obtainable using the correlated 
clustering at the perpendicular direction can be least contaminated by this uncertainty, and 
there will be a way to separately extract this remaining information from other contaminated 
parts along the LOS.
We apply the wedge approach \citep{Kazin_2013,Sanchez_2014,Cris_2016,Ross_2017,Sanchez_2017}
to probe the uncontaminated BAO feature by binning the angular 
direction from the perpendicular to radial directions, and try to successfully recover the 
residual BAO peak that has survived and 
get constraints on $D_{A}$ and $H^{-1}$.
Recovering the real-space correlation function at small scales 
($s<50\,\mathrm{h}^{-1}\mathrm{Mpc}$) has been done recently using 
the deprojection method by \cite{Srivatsan}, but it is of major interest to check the impact 
of this effect on the BAO peak when using photo-z's.

The paper is described as follows: In Section \ref{sec:Section2} we
describe the different correlation functions we calculate 
and introduce the catalogue which we use for the analysis.
In Section \ref{sec:Section3} we describe the theoretical RSD model that we use to get
the correlation function at the targeted redshift and analyse
the BAO peak obtained from \xismu. We also compare the 68\% and 95\% confidence limits
on the fiducial values of $D_{A}$ and $H^{-1}$ obtained from the spectroscopic and photometric 
samples. We summarise all our results in Section \ref{sec:conclusion}.


\section{Remaining BAO feature in photometric map}\label{sec:Section2}
The excess probability of finding two objects relative to a Poisson distribution at volumes 
$dV_{1}$ and $dV_{2}$ separated by a vector distance \textbf{r} is given by the two-point 
correlation function $\xi(r)$~\citep{totsuji_1969,davis_peebles_1983}. The galaxy 
distribution seen in redshift space exhibits an anisotropic feature distorting 
$\xi(r)$ into $\xi(\sigma,\pi)$ along the LOS where $\sigma$ and $\pi$ denote the 
transverse and radial components of the 
separation vector \textbf{r}. Acoustic fluctuations of the 
baryon--radiation plasma of the primordial Universe leaves the signature on the density 
perturbation of baryons. 
This standard ruler length scale, set by the acoustic wave, propagates until 
it is frozen at decoupling epoch to remain in the large scale structure of the Universe. The 
threshold length scale of acoustic wave is called as the sound horizon, which is given by,
\begin{equation} 
r_{s} = \int_{z_{drag}}^{\infty} \frac{c_{s}(z)}{H(z)}dz \end{equation}
where $c_{s}$ is sound speed of the plasma. In a wide deep field spectroscopic galaxy survey, 
the signature of the BAO wave is observed in
precisely determined redshift space, which opens an opportunity to separately access 
transverse and radial cosmic distances. If the target galaxy distribution is given by 
photometrically determined redshift, then it is expected that both measured distances are 
differently contaminated by the uncertainty in redshift determination. We present diverse 
correlation function estimators below, and find an optimal one to probe 
the least contaminated distance measure.

\begin{figure*}
	\includegraphics[width=\textwidth]{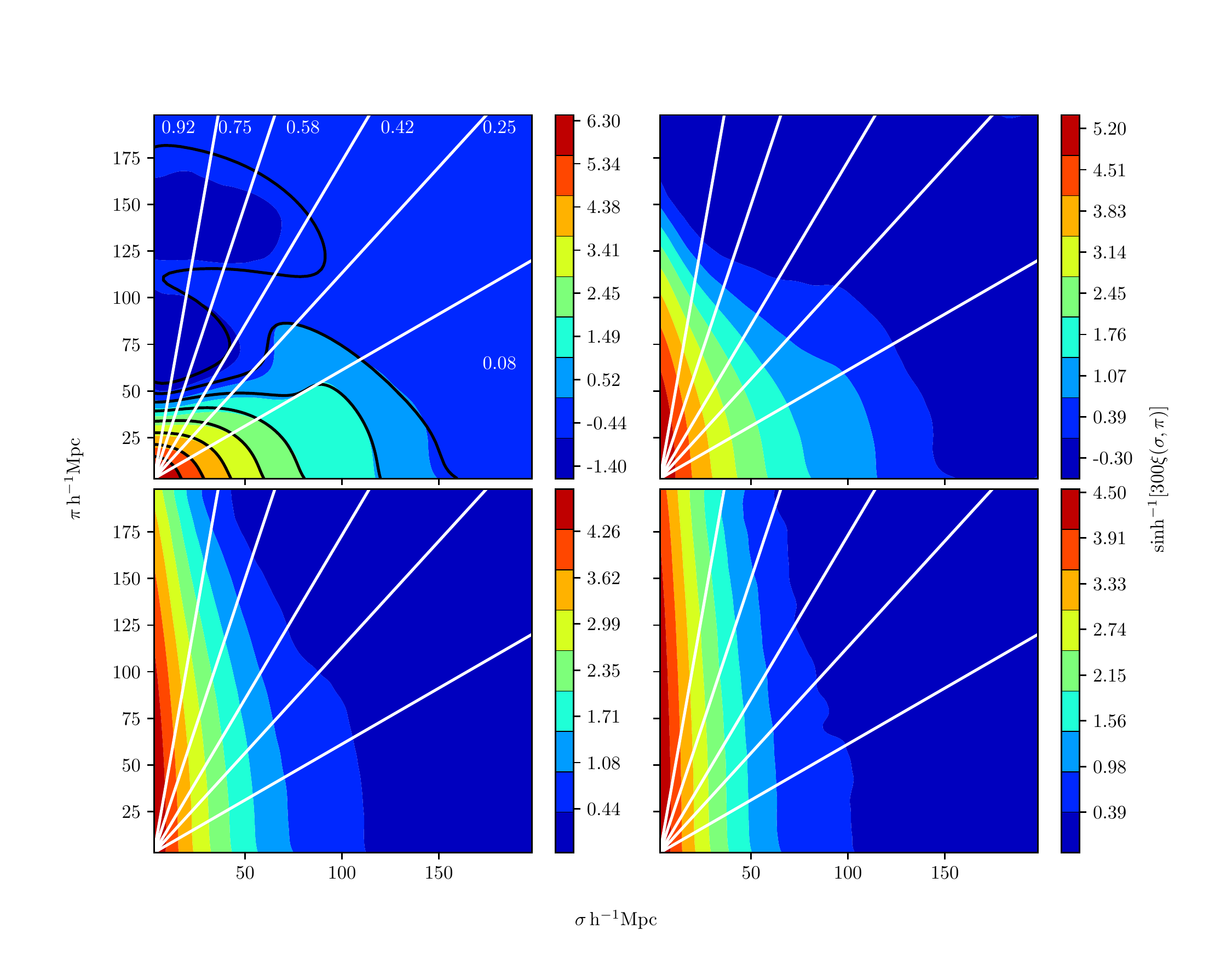}
	\caption{The two-dimensional correlation function for the spectroscopic
		     sample (top left), \sigmaone photometric sample (top right), \sigmatwo photometric
		     sample (bottom left) and \sigmathree photometric sample (bottom right) given by
		     the coloured contours. 
		     The BAO feature which is clearly visible in the spectroscopic 
		     sample ($\sim 100\,\mathrm{h}^{-1}\mathrm{Mpc}$) is smoothed out in the 
		     photometric samples (the smoothing becomes greater with 
		     increasing photometric uncertainty). The function plotted in the colour
		     bar is sinh$^{-1}(300)\xi(\sigma,\pi)$, which is linear near zero, but logarithmic 
		     for high values of $\xi(\sigma,\pi)$. 
		     The solid black line in the top left panel is the 
		     \xisigpi calculated (at the same redshift as the sample) from the theoretical 
		     model explained in Section \ref{sec:theoretical_model}. The solid white lines denote
		     the different $\mu$ bins in which we calculate \xismu and the 
		     mean values of each $\mu$ bin is written in white.
		     Units on both the axes are in $\mathrm{h}^{-1}$Mpc.}
	\label{fig:contour}
\end{figure*}

\subsection{The simulated photometric map} \label{sec:mock_DR12_catalogue}
The photometric galaxy distribution simulations are made using 1000 simulations mocking 
the galaxy distributions and survey geometry of DR12 CMASS catalogue \citep{Manera}.
The base spectroscopy DR12 CMASS simulations are generated using the Quick Particle Mesh 
method (QPM) in which the angular selection function and redshift distribution of 
selected targets in DR12 CMASS are mimicked.
Haloes have been populated with mock galaxies using a calibrated halo occupation distribution 
prescription in those simulations.
The fiducial cosmology used is $\Omega_{m} = 0.274$, $\Omega_{b} = 0.046$, $\Omega_{\Lambda} = 
0.726$, $h = 0.7$, $n = 0.95$ and $\sigma_{8} = 0.8$. 
The original CMASS simulations include both the northern and southern skies, but only 
the northern
sky simulation is used in this manuscript. The given CMASS simulation is provided in the 
redshift range of $0.43<z<0.7$, and only simulated galaxies at $0.53<z<0.63$ are used in this
paper.
The angular positions in the CMASS simulations are used without alterations, with only the 
redshift being altered for mocking the photo-z uncertainty. In reality,  
the statistical nature of the photo-z error is more complicated to be 
specified with any known distribution function, but it is assumed that the error propagation of
photo-z uncertainty into cosmological information is mainly caused by the dispersion length. 
Thus the simple Gaussian function of statistical distribution is chosen for photo-z uncertainty
distribution, and we apply the various photo-z error dispersion $\sigma_z$ which is given by,
\begin{equation}\label{eqn:sigma_z}
\sigma_{z} = \sigma_{0}\times(1+z_{spec}) ,
\end{equation}
where $\sigma_{0}$ denotes the photo-z uncertainty dispersion at $z=0$. In reality, the 
precision is dependent on many factors such as magnitude and spectral type, but here only 
the redshift factor is counted in Eq.~\ref{eqn:sigma_z} in which the coherent 
statistical property 
determined only by $z$ is applied for all types of galaxies in the simulation.

The spectroscopically determined redshift precision is expected to be $\sigma_{z}/(1+z) \sim 
0.0005$ \citep{DESI_1}
in which the coherent length scale is much bigger than the physical length difference 
caused by photo-z uncertainty. Thus the given redshifts of the CMASS simulations are assumed 
to be determined by spectroscopy.
Then the generic photometric redshifts are assigned to each galaxy by random extraction from a 
Gaussian distribution with mean equal to the galaxy spectroscopic redshift and standard 
deviation equal to the assumed photometric redshift error of the sample. 
Certainly, a photometric survey will provide us with more galaxies observed, which reduces 
the shot noise to improve the accessibility towards smaller scale clustering. But in this 
verification work, note that the total number of targeted galaxies are the same for both  the
photometric and spectroscopic samples. We focus on the cosmological information loss caused by 
photo-z uncertainty, without considering the benefit of more galaxy samples in a photometric 
survey.

The photometric $z$ determination error for ongoing or planned surveys is estimated to be 
around $0.01<\sigma_{0}<0.03$ \citep{DES,euclid_red_book,Ascaso}. The error on the 
photometric redshift obtained from a Luminous Red Galaxy (LRG) sample from the recent 
DECaLS DR7 \citep{Decals} data (covering part of the DESI footprint) by 
Zhou. et al (2019, in preparation) is $0.02<\sigma_{0}<0.03$. Thus it is reasonable to test
the error propagation with the selected $\sigma_{0}$ as $\sigma_{0} = (0.01,0.02,0.03)$ which 
ranges from the optimistic to the conservative estimations from the surveys. We present the 
galaxy distribution showing the LOS positional dislocation in Fig.\ref{fig:Y_vs_Z_plot}, in 
which Y and Z denote the tangential and radial directions. The simulated galaxy 
distribution is shown at the top panel, and the dislocated galaxy distributions with 
$\sigma_{0} = 0.01$, $0.02$ and $0.03$ are presented from the second to the bottom panels 
respectively. The change of galaxy distribution is visible in those panels.

\begin{figure*}
	\includegraphics[width=\columnwidth]{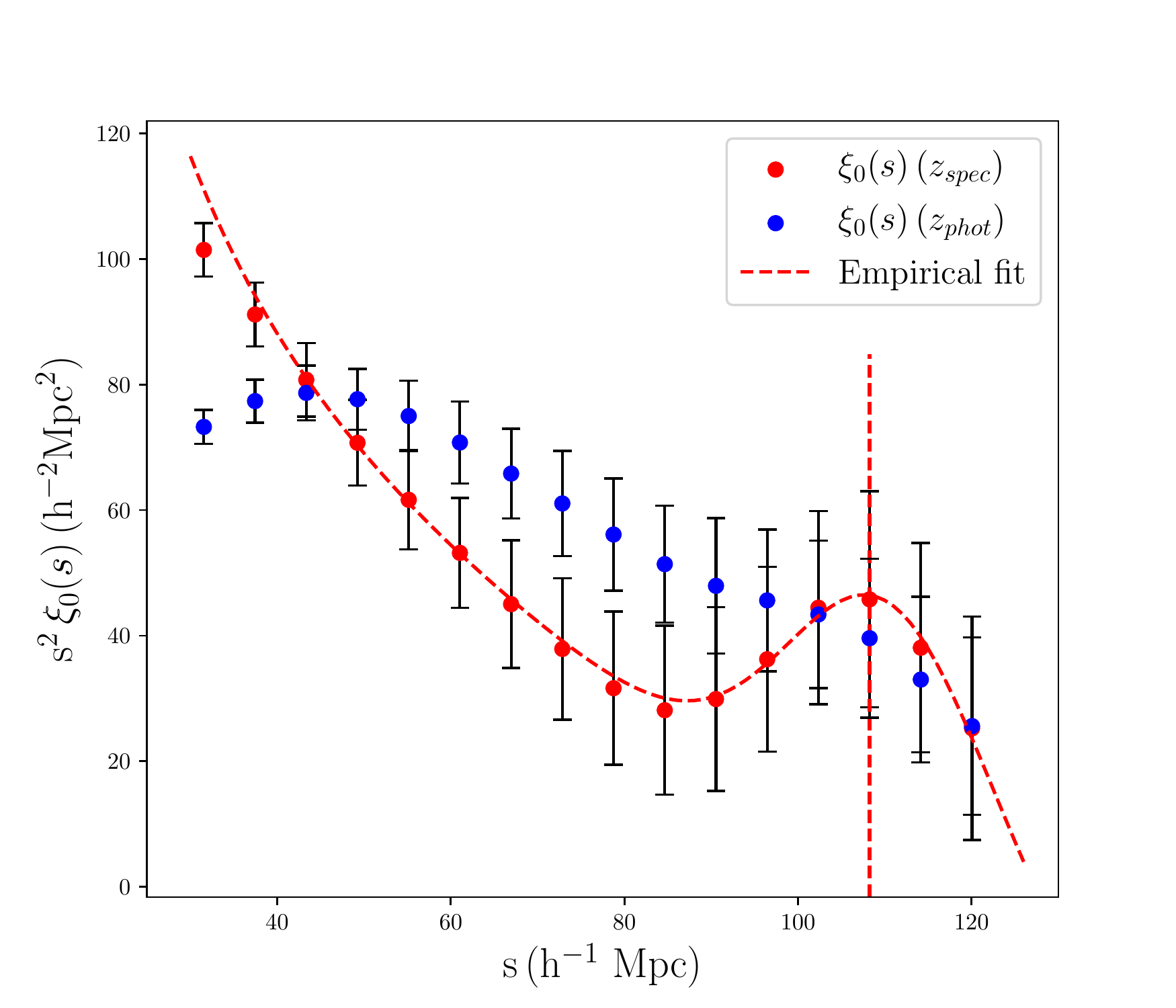}
	\includegraphics[width=\columnwidth]{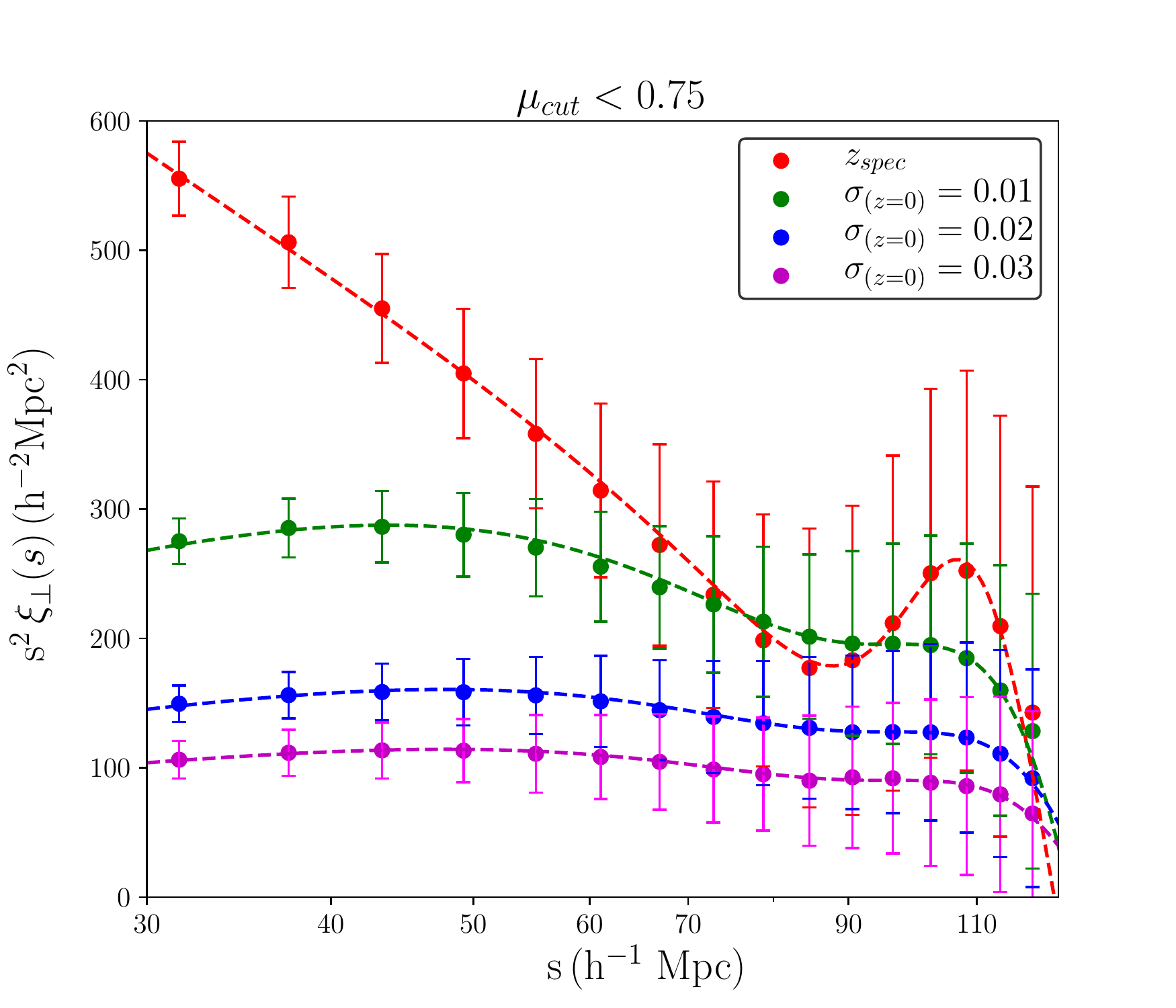}   
	\caption{{\it Left panel:} The angle averaged monopole correlation function 
	         (multiplied by $s^{2}$) calculated 
		     for the spectroscopic sample (plotted in red) and for the \sigmaone 
		     photometric sample (plotted in blue) within the range $0.53<z<0.63$. 
		     The red dotted line shows the best-fit obtained from the empirical model in 
		     Eq.~\ref{eqn:empirical_fit} for the spectroscopic sample.
		     The vertical red dotted line shows the $s_{m}$ obtained from the best-fit
		     to the spectroscopic sample. 
		     {\it Right panel:} The projected correlation function \xiproj (multiplied by 
		     $s^{2}$) defined by 
		     Eq.~\ref{eqn:proj_xismu} with $\mu_{\mathrm{cut}}<0.75$ 
		     for the spectroscopic sample (in red), 
		     the \sigmaone (in green), \sigmatwo (in blue) and \sigmathree (in magenta) 
		     photometric samples. The dotted 
		     lines represent the best-fit model to the data obtained from 
		     Eq.~\ref{eqn:empirical_fit}. The error bars plotted in both the panels
		     have been calculated using the full covariance matrix 
		     as mentioned in Eq.~\ref{eqn:cov_matrix}.} 
	\label{fig:monopole_and_3PROJECTED}   
\end{figure*}


\subsection{The BAO peaks imprinted on diverse correlation functions}
\label{sec:correlation_function}

The BAO feature is imprinted on the correlation function through the integrated effect of BAO 
signatures that remain in the power spectrum. We introduce all different configurations to 
describe the correlation function in redshift space, and discuss the optimised correlation 
function configuration to probe the BAO peaks from the photometric map. The correlation 
function $\xi$ is estimated using the Landy \& Szalay estimator (hereafter LS) which is known 
to be less sensitive to the size of the random catalogue and also handles edge corrections 
better~\citep{kerscher}. The LS estimator in $(s,\mu)$ coordinates is given by,
\begin{equation}\label{eqn:landy_szalay_smu}
\xi(s,\mu) = \frac{DD(s,\mu) - 2DR(s,\mu) + RR(s,\mu)}{RR(s,\mu)} ,
\end{equation}
where $DD$, $DR$ and $RR$ refer respectively to the number of data-data pairs, data-random 
pairs and the random-random pairs within a spherical shell of radius $s$ and $s+ds$ and the 
angle to the LOS $\mu$ and $\mu+d\mu$. 
The radius to shell $s$ and the observed cosine of the angle the pair makes with respect to 
the LOS $\mu$ are given by
$s^2=\sigma^2+\pi^2$ and $\mu=\pi/s$ respectively, where $\sigma$ and $\pi$ denote the 
transverse and radial directions. 

It is common practice to separate the random sample distributions into the angular and redshift
components	separately. For the angular components, we create random objects within the RA and 
DEC limits of the data catalogue. The number of objects are usually twice or more than the data
catalogue to avoid shot noise effects. In our case the random catalogue has 5 times more 
objects than the data catalogue. For the redshift component, from the data catalogue we extract
redshifts randomly within the chosen redshift 
range \citep[see][for more info]{Ross_2012,Veropalumbo_2016}. 
We use the publicly available {\tt KSTAT} (KD-tree Statistics Package) code \citep{KSTAT} to 
calculate all our correlation functions. A separate random catalogue is created for each 
realisation of the mocks because each realisation has it's own distinct redshift distribution. 
We have verified that the difference between the correlation function calculated using the 
single spectroscopic random catalogue provided and the correlation function calculated using 
the distinct random catalogues is of the order of 3-5\%. 
	
In Fig.\ref{fig:contour}, the anisotropy correlation function distorted in redshift space 
is presented in cartesian coordinates. The spectroscopic sample is given in the top-left 
panel along with the three different photometric samples in the other panels as denoted 
in the figure. We use a sinh$^{-1}(300)\xi(\sigma,\pi)$ transform for the colour 
bar representation for easy visualisation as $\xi(\sigma,\pi)$ values extend from small to 
large values. The sinh$^{-1}(x)$ function equals $x$ for $x\ll 1$ and ln $x$ for $x\gg 1$. 
BAO features are clearly observed at the outer contour encompassing 
$s\sim 110\,\mathrm{h}^{-1}\mathrm{Mpc}$ at both transverse and radial directions for the 
spectroscopy case, but are smeared out by the uncertainty of redshift measurement along 
the LOS for all the photometric cases. However, the errors propagate differently depending 
on the directions. It is not clearly visible if there are any remaining BAO features for 
the photometric cases. Thus we explore other diverse correlation configurations to extract 
the remaining BAO feature below.

When the objects have a spectroscopic redshift, it is useful to calculate the monopole 
correlation function $\xi_{0}(s)$, which is obtained by integrating $\xi(s,\mu)$ in $\mu$ 
direction,
\begin{equation}\label{eqn:mono_xismu}
\xi_0(s) = \int^1_0 d\mu'W(\mu':\mu_{\rm cut}=1)\xi(s,\mu') ,
\end{equation}
where the weighting function $W(\mu':\mu_{\rm cut}=1)$ is given by $W(\mu':\mu=0$) at 
$\mu'>\mu$, and it is normalised as, 
\begin{equation}
\int^1_0 d\mu'W(\mu':\mu_{\rm cut}) = 1 .
\end{equation}
The cut--off $\mu$ is set to be $\mu_{\rm cut}=1$ for the monopole correlation function. 
In the left panel of Fig.\ref{fig:monopole_and_3PROJECTED}, BAO features observed 
from $\xi_{0}(s)$ using both the spectroscopic and photometric simulation maps are 
represented by red and blue points are respectively.
While the peak is certainly visible around $s\sim 110\,\mathrm{h}^{-1}\mathrm{Mpc}$ in the 
spectroscopic map, it is smoothed out in the photometric map 
due to the uncertainty in the radial distance determination, with only the power law 
shape remaining at scales $s<80\,\mathrm{h}^{-1}\mathrm{Mpc}$ \citep{Farrow,Srivatsan} 

As most contaminated pairs are found along the
radial configuration, the correlation pairs at higher $\mu$ is trimmed out, and the cutoff 
$\mu$ is redefined in Eq.~\ref{eqn:mono_xismu} as $\mu_{\rm cut}<1$.
This incompletely integrated correlation function $\xi_{\perp}$ with the non--trivial 
$\mu_{\rm cut}$ will be an alternative option dubbed as the projected correlation function
and is given by,
\begin{equation}\label{eqn:proj_xismu}
\xi_{\perp}(s) = \int^1_0 d\mu'W(\mu':\mu_{\rm cut}<1)\xi(s,\mu')\,.
\end{equation}
The reconstructed BAO features are obtained by trimming out the contaminated configuration 
along the LOS, in which a commonly used value of $\mu_{\rm cut}=0.75$ \citep{Ross_2017} 
is applied. The results are presented in the right panel of 
Fig.~\ref{fig:monopole_and_3PROJECTED} for the spectroscopy and three photo-z 
uncertainty cases of $\sigma_0=(0.01,0.02,0.03)$.
Although the observed BAO features from the photometric maps aren't as clearly visible as 
the spectroscopy case, the shape of the correlation function is visibly improved using
the projected correlation function.

\begin{figure*}
	\includegraphics[width=\textwidth]{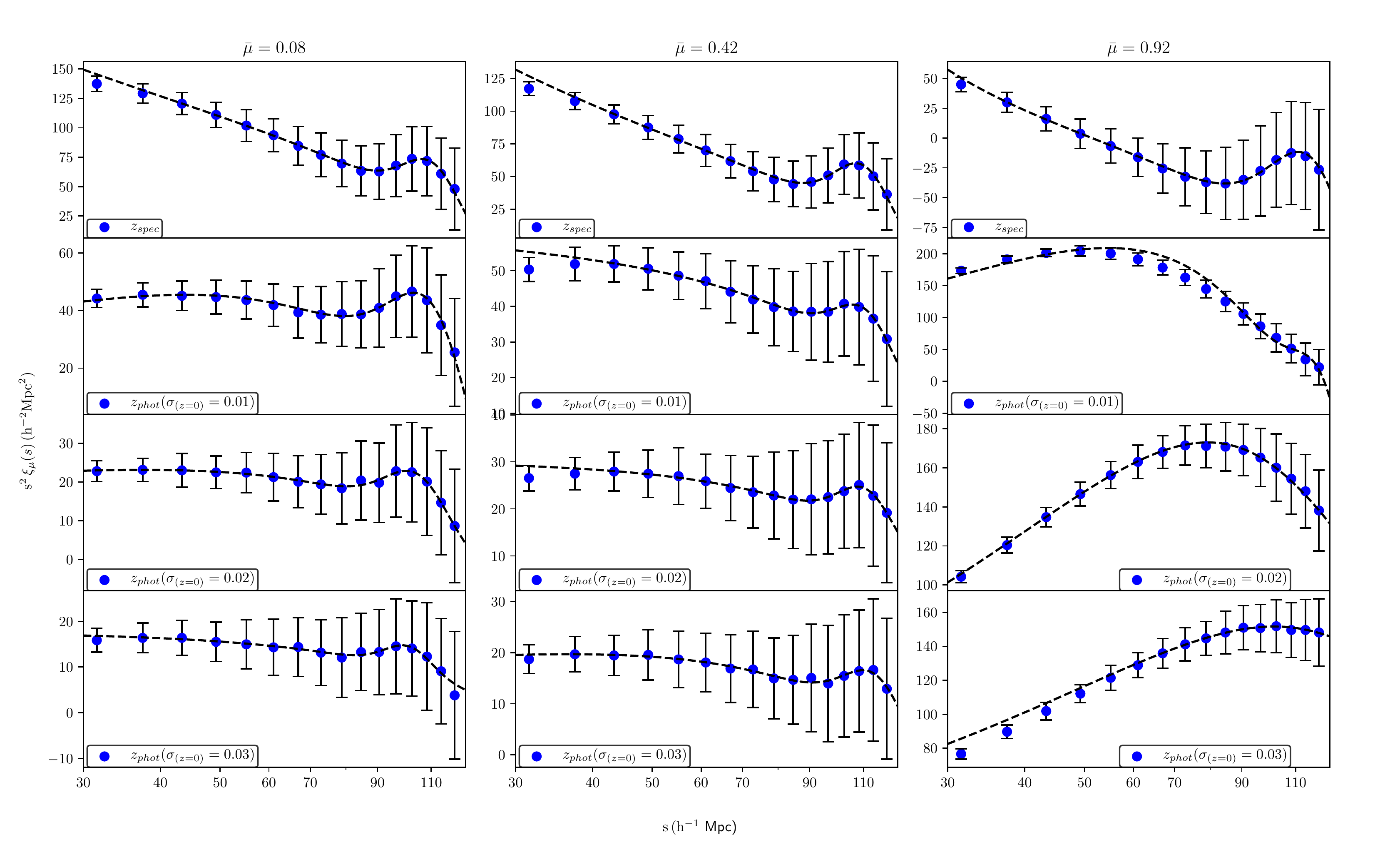}  
	\caption{The correlation function \xismu (multiplied by $s^{2}$) 
		     calculated by splitting into wedges of $\mu$ given 
		     by the blue dots. The dashed black lines in each plot 
		     show the best-fit obtained from the 
		     empirical model in Eq.~\ref{eqn:empirical_fit}.
		     The first, second and 
		     third columns in the figure represent $\bar \mu=0.08$, 0.42 and 0.92 bin 
		     respectively and the first, second, third and fourth rows in the figure 
		     represent the spectroscopic ($\sigma_{0}\approxeq 0$), \sigmaone, 
		     \sigmatwo and \sigmathree photometric samples respectively. The error bars 
		     plotted have been calculated using the full covariance matrix as mentioned in 
		     Eq.~\ref{eqn:cov_matrix}.} 
	\label{fig:s_vs_xismu_zspec_vs_zphot_all}   
\end{figure*}

The improvement by applying the projected correlation function suggests that the 
contaminated pairs can be removed by sorting the correlation function in $\mu$ bins. 
We pay attention to the usefulness of exploiting the wedge correlation function to 
separate the radial contamination from the BAO signal imprinted on perpendicular 
configuration pairs. The wedge correlation function $\xi_w$ is given by,
\begin{equation}\label{eqn:wed_xismu}
\xi_w(s,\mu_i) = \int^1_{\mu_i^{\rm min}} d\mu'W(\mu':\mu_{\rm cut}=\mu_i^{\rm max})\xi(s,\mu') ,  
\end{equation}
where $\mu_i$ is the mean $\mu$ in each bin, and $\mu_i^{\rm min}$ and $\mu_i^{\rm max}$ 
are the minimum and maximum values of $\mu$. We choose 6 bins in the $\mu$ direction 
with $\Delta \mu=0.17$ between $\mu=0$ and 1. The wedge correlation functions at $i=1,3,6$ 
are presented at the left, middle and right panels in
Fig.~\ref{fig:s_vs_xismu_zspec_vs_zphot_all}. 
The $\xi_w$ with diverse photometric errors of $\sigma_0=(0,0.01,0.02,0.03)$ are shown from 
the top to the bottom panels. The BAO features are more contaminated at higher $i$.

\subsection{Measurement of the residual BAO peaks} 

The empirical model that we use to fit the correlation function and obtain the BAO peak 
location is the one proposed by \citet{Sanchez_empirical}, which is used to interpolate the 
correlation function at the BAO scales. It is given by:	
\begin{equation}\label{eqn:empirical_fit}
\xi_{\rm mod}(s) = B + \left( \frac{s}{s_{0}} \right) ^{- \gamma} + \frac{N}{\sqrt{2\pi \sigma^{2}}} \mathrm{exp} \left( -\frac{(s-s_{m})^{2}}{2\sigma^{2}} \right) ,
\end{equation}
where $B$ takes into account a possible negative correlation at very large scales, 
$s_{0}$ is the correlation length (the scale at which the correlation function $\simeq$ 1) 
and $\gamma$ denotes the slope. These 3 parameters model the correlation function at small 
scales of $s < 50 \mathrm{h}^{-1}$Mpc, and are fixed firstly by fitting the data at scales below the 
BAO peak, i.e. $30 <s\,(\mathrm{h}^{-1}\mathrm{Mpc}) < 80$. The remaining three parameters, 
$N$, $\sigma$ and $s_{m}$ are the parameters of the Gaussian function that model the BAO 
feature and, in particular, $s_{m}$ represents the estimate of the BAO peak position. This 
empirical model can be used to accurately predict the BAO peak 
position~\citep{Veropalumbo_2016} when the correlation function is provided.

\begin{figure*}
\includegraphics[width=\columnwidth]{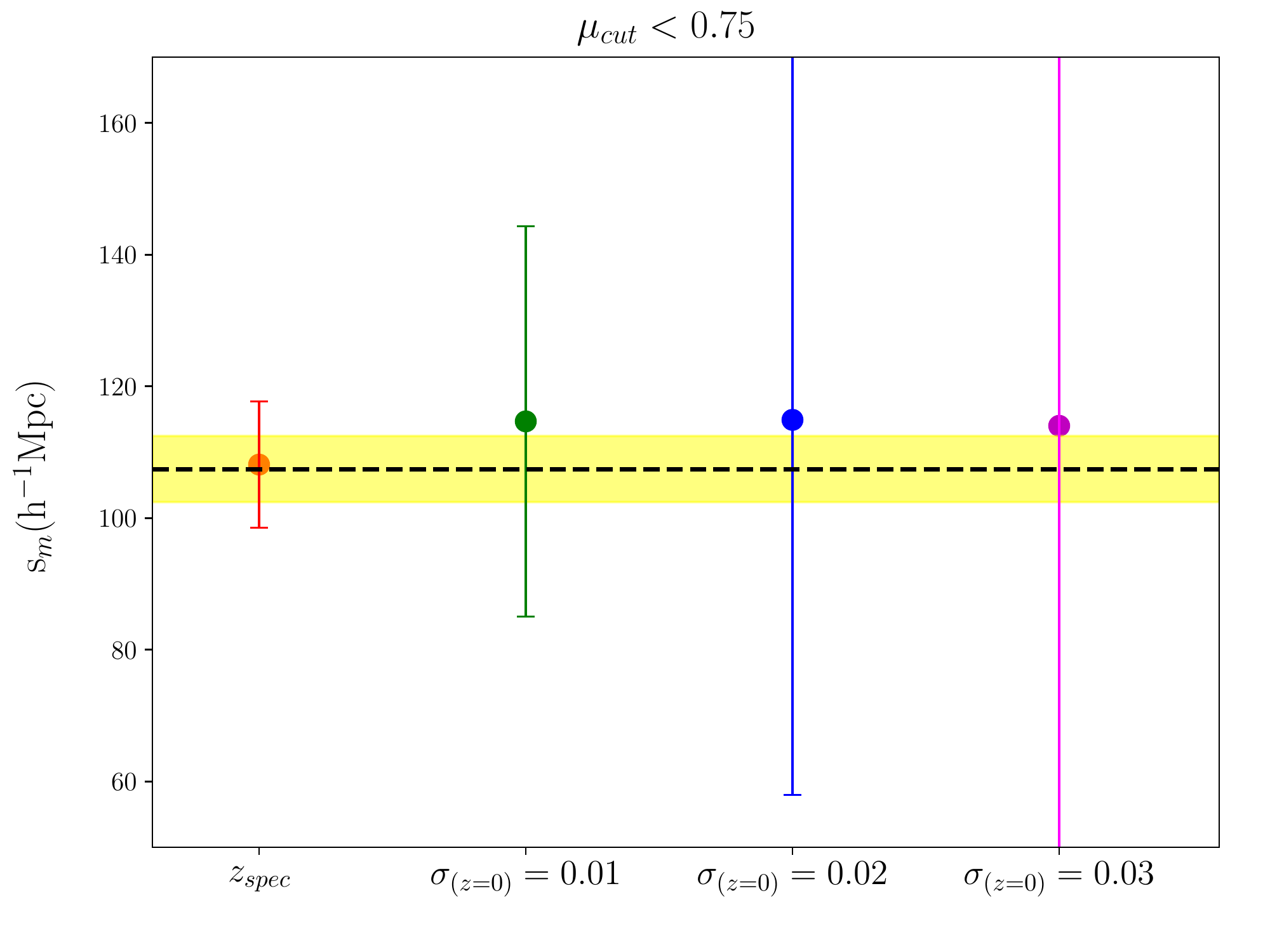} 
	\includegraphics[width=\columnwidth]{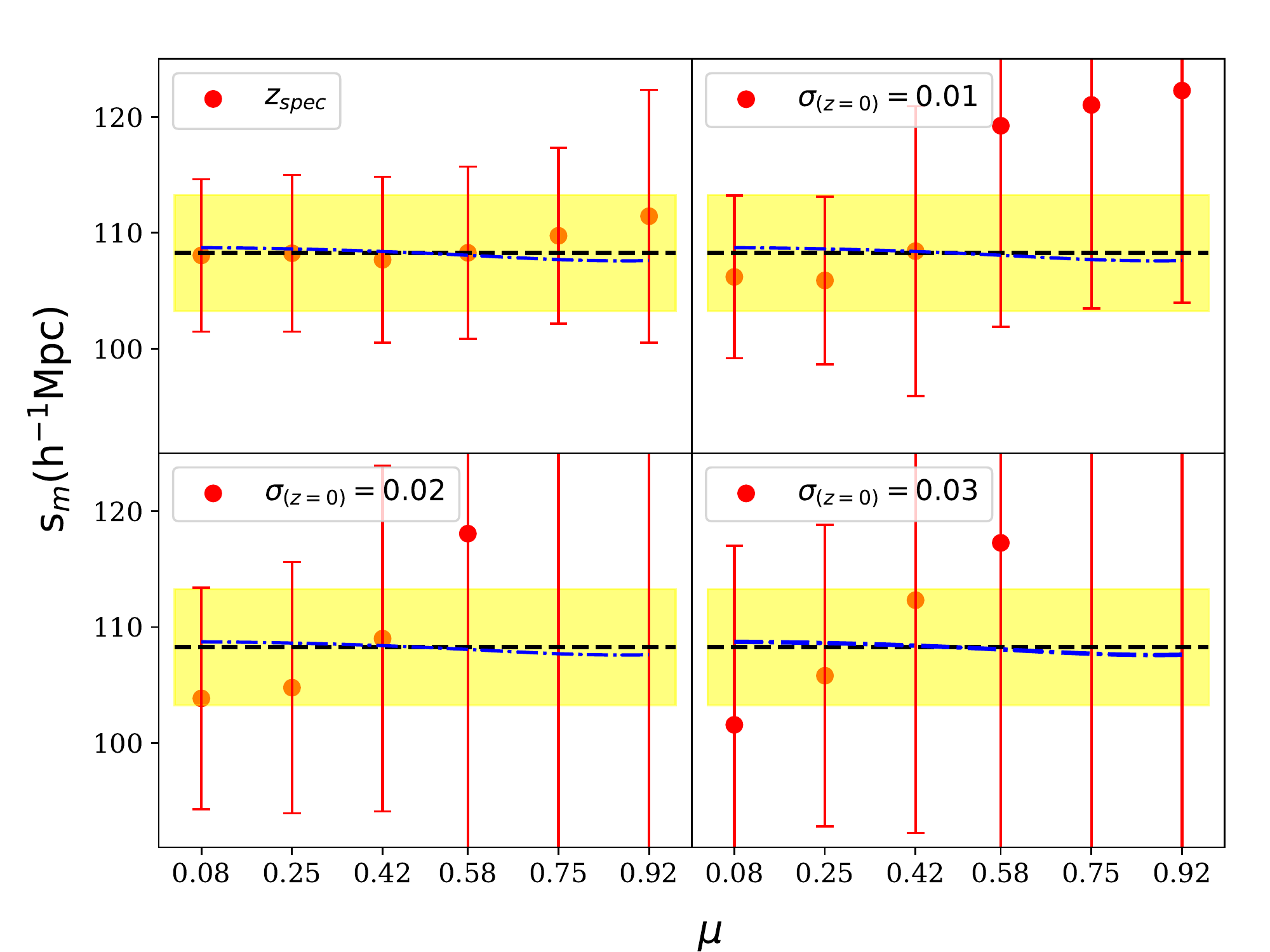}   
	\caption{\textit{Left panel:} The values of $s_{m}$ obtained from \xiproj with 
	         $\mu_{cut}<0.75$. 
		     The x-axis denotes the spectroscopic, \sigmaone, \sigmatwo and \sigmathree
		     photometric samples as marked by the labels. 
		     The black dotted line in both the panels represents the value of $s_{m}$ 
		     obtained from the empirical fit for $\xi_{0}(s)$ calculated on the 
		     spectroscopic sample and the error on the same is given by the yellow
		     highlighted region.
		     \textit{Right panel:} 
		     The x-axis denotes the 6 $\mu$ bins we have used and the y-axis denotes the 
		     value of $s_{m}$ obtained from the empirical fit for the spectroscopic sample
		     (top left), \sigmaone (top right), \sigmatwo (bottom left) and \sigmathree
		     (bottom right) photometric samples. 
		     The blue dash-dotted line represents the $s_{m}$ obtained 
		     for the different $\mu$ bins from the theoretical template.}
	\label{fig:mu_vs_BAO_peak}   
\end{figure*}

The means of the projected and wedge correlation functions are measured using 200 realisations. 
The covariance matrix of $\xi_{\perp}(s)$ and $\xi_w(s,\mu_i)$ is computed as,
\begin{equation}\label{eqn:cov_matrix}
C^{ij}_{X}(\xi^i_{X},\xi^j_{X})=\frac{1}{N-1}\sum^{N}_{n=1}[\xi^n_{X}(\vec{x}_i)-\overline{\xi}_{X}(\vec{x}_i)][\xi^n_{X}(\vec{x}_j)-\overline{\xi}_{X}(\vec{x}_j)],
\end{equation}
where $X$ denotes the symbols of $\perp$ or $w$ representing the projected or wedge 
correlation functions respectively, and the total number of $N$ is given by $N=200$. 
The $\xi^n_{X}(\vec{x}_i)$ represents the value of the projected or wedge correlation 
function of $i^{th}$ bin of $\vec{x}_i$ in the $n^{th}$ realisation, and 
$\overline{\xi}_{X}(\vec{x}_i)$ is the mean value of $\xi^n_{X}(\vec{x}_i)$ over all the 
realisations. The number of realisations exceeds the number of $(s,\mu)$ bins of 96 bins, 
and the inverse of $C_{ij}$ is well defined and thus does not require any de-noising 
procedures such as singular value decomposition. Additionally, we also count the offset 
caused by the finite number of realisation as, 
\begin{equation}
C^{-1}=\frac{N_{mocks} - N_{bins}-2}{N_{mocks}-1}\ \hat{C}^{-1}\ ,
\end{equation}
where $N_{bins}$ denotes the total number of $i$ bins.

The fitting parameter space is given by $x_p=(B,s_0,\gamma,N,s_m,\sigma)$, and the BAO peak 
$s_m$ for the projected correlation function is estimated after fully marginalising all 
other parameters in $x_p$. The fitting function is given by,
\begin{equation}
\chi^2(x_p)=\sum_{s,s'}(\xi_{\rm mod}(s)-\xi_{\perp}(s))C^{-1}_{s;s'}(\xi_{\rm mod}(s')-\xi_{\perp}(s'))
\end{equation}
where $C^{-1}_{s;s'}$ denotes the inverse covariance matrix for two different separation 
distances $s$ and $s'$. For the wedge correlation function, the $\chi^2_{\mu_i}$ at each 
$\mu_i$ bin is expressed as,
\begin{equation}
\chi^2_{\mu_i}(x_p)= \sum_{s,s'}(\xi_{\rm mod}(s)-\xi_{w}(s,\mu_i))C^{-1}_{s;s'}(\mu_i)(\xi_{\rm mod}(s')-\xi_{w}(s,\mu_i))
\end{equation}
where $C^{-1}_{s;s'}(\mu_i)$ is the sub--inverse covariance matrix of $C^{-1}$ including the 
$\mu_i$ coordinate.

For $\xi_{\perp}$ with a non-trivial cut of $\mu_{\rm cut}<0.75$, Eq.\ref{eqn:empirical_fit}
is used to get $s_{m}$ for all the samples with varying $\sigma_{0}$. The $s_{m}$ values 
for all the samples are plotted in the left panel of Fig.\ref{fig:mu_vs_BAO_peak}. For 
the spectroscopic sample, $s_{m}$ is measured with a fractional error of 8\%, but the error 
propagation into the clustering by the photo-z uncertainty increases towards the radial 
direction. So, even though the $s_{m}$ obtained for the photo-z samples seem to be accurate
within $1\sigma$ compared to the spectroscopic case, they are not precise, i.e. the errors 
are large. The error on $s_{m}$ increases with increasing $\sigma_{0}$. 

Thus, to clearly examine the contamination of the pairs along the radial direction, it is 
necessary to split the correlation function into smaller $\mu$ bins and measure 
$\xi_{w}(s,\mu_i)$. The $s_{m}$ values obtained from $\xi_{w}(s,\mu_i)$ for the four 
$\sigma_{0}$ samples in the 6 $\mu$ bins used is shown in the right panel of 
Fig.\ref{fig:mu_vs_BAO_peak}. 
As a benchmark, we make use of the $s_{m}$ obtained from 
$\xi_{0}(s)$ using the spectroscopic sample (black dotted line, with $1\sigma$ error given 
by the yellow highlighted region) along with the $s_{m}$ obtained from the theoretical 
template (explained in Section \ref{sec:theoretical_model} and given by the blue dot-dashed
line). For all $\bar{\mu}<0.42$, it can be seen that the $s_{m}$
obtained is within $1\sigma$ compared to the $s_{m}$ from $\xi_{0}(s)$. On the other hand, 
for all $\bar{\mu}>0.42$, only the $s_{m}$ from the spectroscopic sample seems to be within 
$1\sigma$. 
This strongly suggests that measuring the BAO peak within $\bar{\mu}<0.42$ gives
us tight constraints even when we have photometric samples with an error of \sigmathree.


\section{The measured cosmic distances using photometric samples}\label{sec:Section3}

The volume distance $D_{V}(z) = \left[(1+z)^{2}D_{A}(z)^{2}cz/H(z)\right]^{1/3}$ is 
measured through the BAO by exploiting the monopole correlation function. While the 
monopole correlation function is preferably used without any concerns regarding the 
computation of the covariance matrix, both the transverse and radial cosmic distances 
can be separately measured using the 2D anisotropy correlation function. The full 
covariance matrix determination for the 2D anisotropy correlation function is much more 
difficult, but the estimated covariance matrix appears to be stable, at least numerically 
with the number of realisations we have used for the simulations. When both the transverse 
and radial distance measures of redshift are precisely probed, both $D_A$ and $H^{-1}$ are 
determined with high precision. In case there exists a systematic uncertainty in determining 
the radial component, the methodology of using the 2D anisotropy correlation function can be 
useful to remove the contaminated cosmological information along the LOS. In this section, we 
verify whether the transverse cosmic distance can be measured in precision regardless of the 
photo-z uncertainty.

\subsection{Theoretical model to fit cosmic distances}
\label{sec:theoretical_model}

\begin{figure*}
\includegraphics[width=\columnwidth]{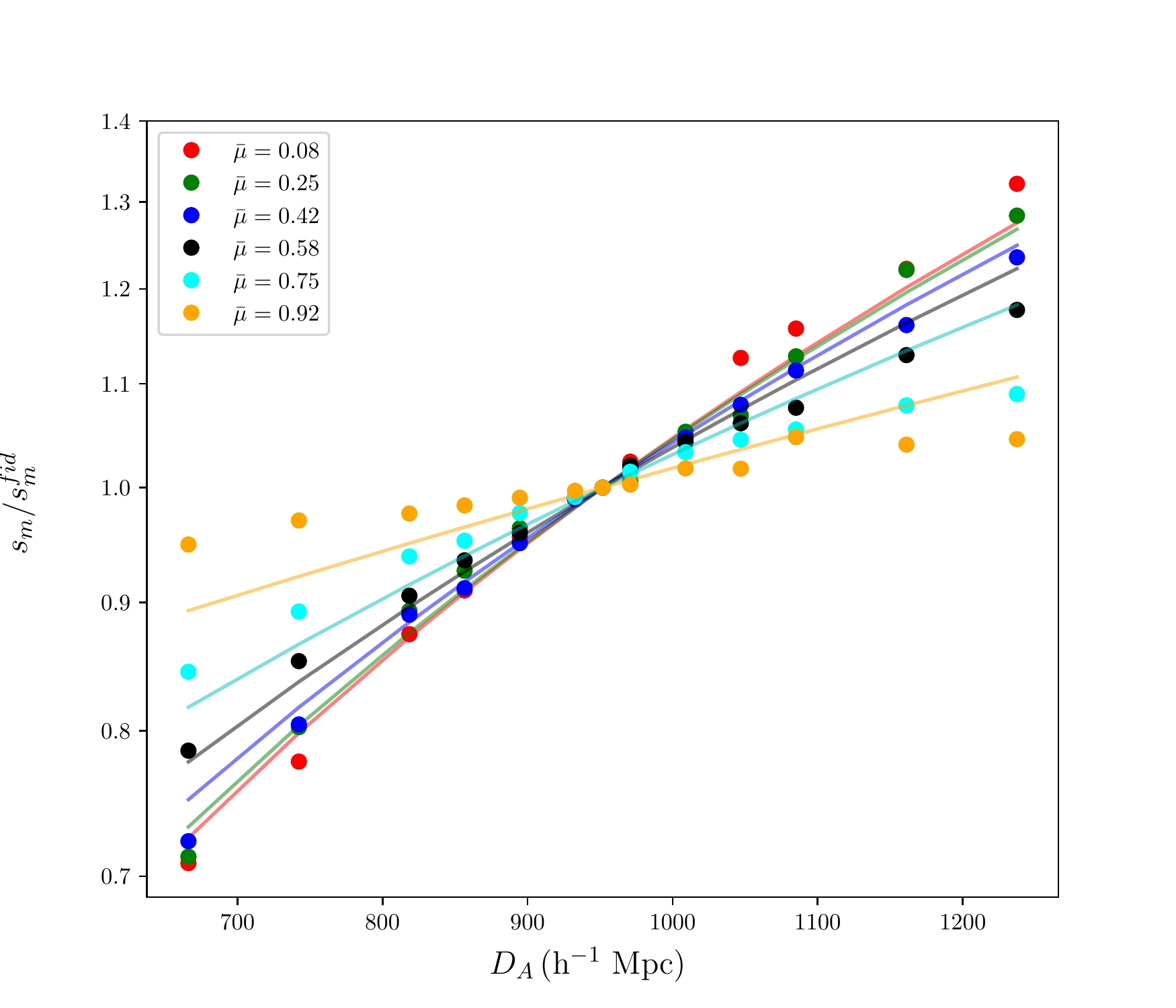} 
\includegraphics[width=\columnwidth]{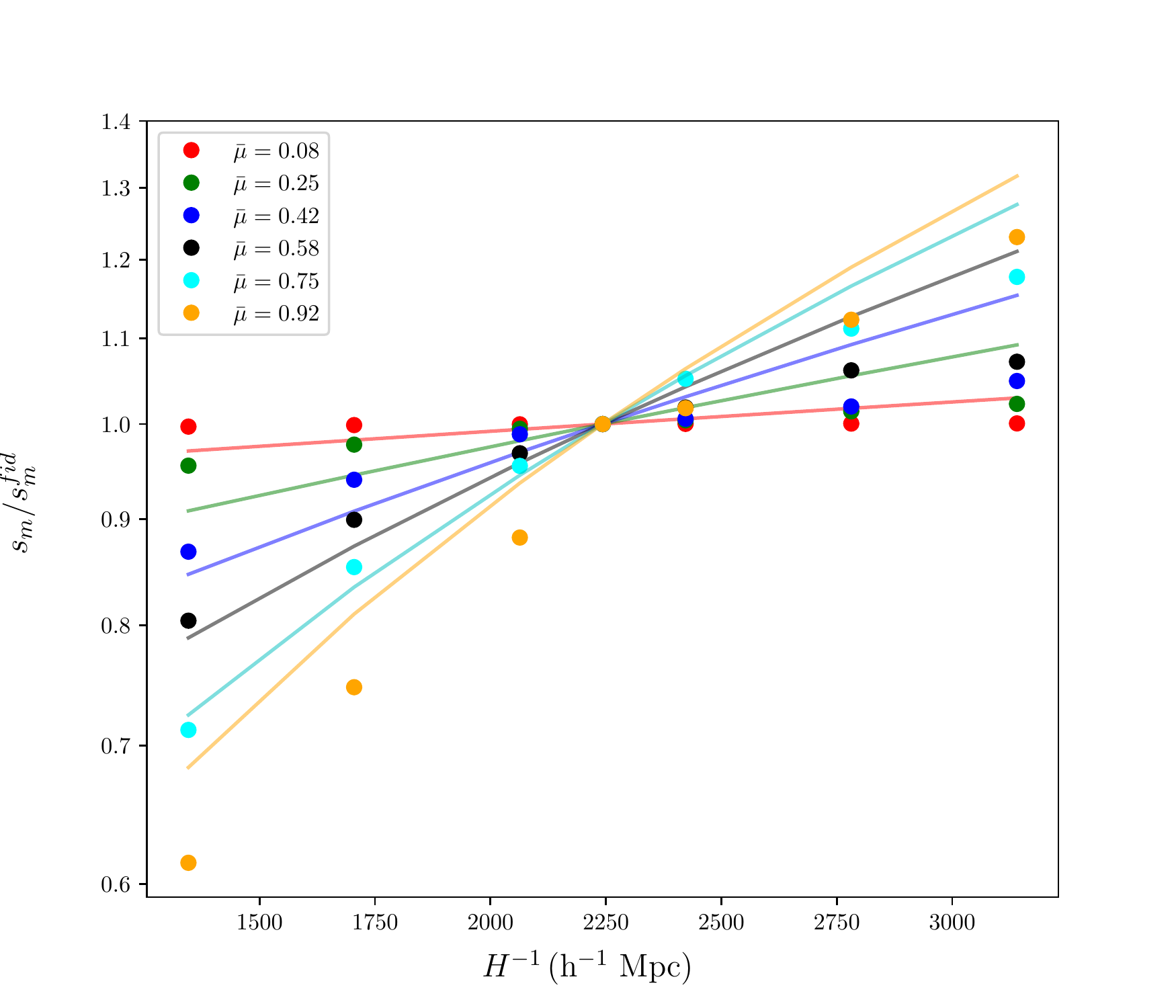}   
	\caption{\textit{Left panel:} The values of $s_{m}/s_{m}^{fid}$ for the 6 $\mu$ bins
	         at different values of $D_{A}$ computed using the TNS model given by the dotted
	         points as colour coded. The solid curves represent the same obtained using a 
	         simple coordinate projection from the fiducial value. 
	         \textit{Right panel:} The same as the left panel, but for different values of 
	         $H^{-1}$. Units are in $\rm \mathrm{h}^{-1} \rm Mpc$.}
	\label{fig:DA_Hinv_var}   
\end{figure*}

We need to theoretically model the correlation function to fit the cosmological distance 
variations. The theoretical correlation function in redshift space $\xi_{\rm th}(s,\mu)$ is 
computed using the improved power spectrum in the perturbative expansion as,
\ba\label{eq:xi_eq}
\xi_{\rm th}(s,\mu)&=&\int \frac{d^3k}{(2\pi)^3} \tilde{P}(k,\mu')e^{i{\bf k}\cdot{\bf s}}\nn\\
&=&\sum_{\ell:{\rm even}}\xi_\ell(s) {\cal P}_\ell(\mu)\,,
\ea
with ${\cal P}$ being the Legendre polynomials. Here, we define $\nu=\pi/s$ and 
$s=(\sigma^2+\pi^2)^{1/2}$. The moments of the correlation function, $\xi_\ell(s)$, are defined by,
\ba
\xi_\ell(s)=i^\ell\int\frac{k^2dk}{2\pi^2}\,\tilde{P}_\ell(k)\,j_\ell(ks)\,.
\ea
The multipole power spectra $\tilde{P}_\ell(k)$ are explicitly given by,
\ba
\tilde P_0(k)&=&p_0(k),\nn\\
\tilde P_2(k)&=&\frac{5}{2}\left[3p_1(k)-p_0(k) \right],\nn\\
\tilde P_4(k)&=&\frac{9}{8}\left[35p_2(k)-30p_1(k)+3p_0(k)\right],\nn\\
\ea
where we define the function $p_m(k)$: 
\ba
p_m(k)&=&\frac{1}{2}\sum_{n=0}^4\frac{\gamma(m+n+1/2,\kappa)}{\kappa^{m+n+1/2}}\,Q_{2n}(k)
\ea
with $\kappa=k^2\sigma_p^2$. The function $\gamma$ is the incomplete gamma function of the 
first kind: 
\ba
\gamma(n,\kappa)=\int^{\kappa}_0dt\, t^{n-1}\,e^{-t}\,.
\ea
The $Q_{2n}$ is explained below.

The observed power spectrum in redshift space $\tilde{P}(k,\mu)$ is written in the following 
form;
\ba
\label{eq:pkred_in_Q}
\tilde{P}(k,\mu) =\sum_{n=0}^8\,Q_{2n}(k)\mu^{2n}\,G^{\rm FoG}(k\mu\sigma_p)\,,
\ea
where the velocity dispersion $\sigma_p$ is set to be a free parameter for FoG effect, 
and the function $Q_{2n}$ are given by,
\ba\label{eq:Q}
Q_0(k)&=&P_{\delta\delta}(k) ,\nn\\
Q_2(k)&=&2P_{\delta\Theta}(k) + C_2(k),\nn\\
Q_4(k)&=&P_{\Theta\Theta}(k) + C_4(k),
\ea
where $C_n$ includes the nonlinear correction terms $A$ and $B$, and $P_{XY}(k)$ denotes the 
power spectrum in real space. The standard perturbation model exhibits the ill-behaved 
expansion leading to the bad UV behaviour which is regularised by introducing UV cut-off in 
this manuscript. The treatment of resummed perturbation theory dubbed as {\tt RegPT} is well 
explained in~\cite{Taruya:2012ut}. The auto and cross spectra of $P_{XY}(k)$ are computed up 
to first order, and higher order polynomials $A$ and $B$ are computed up to zeroth order, 
which are consistent in the perturbative order.

\begin{figure*}
	\includegraphics[width=3.4in]{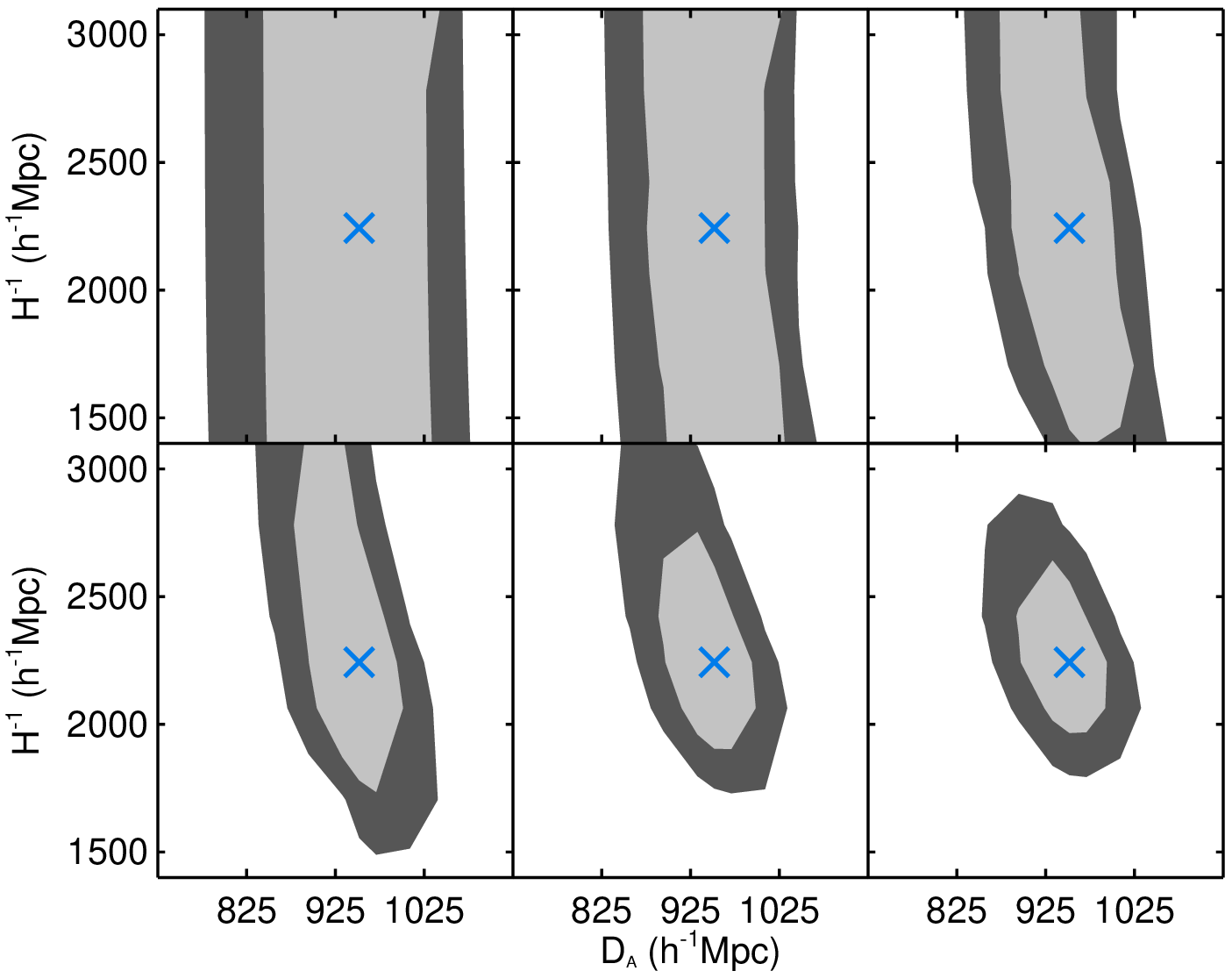}  
	\includegraphics[width=3.4in]{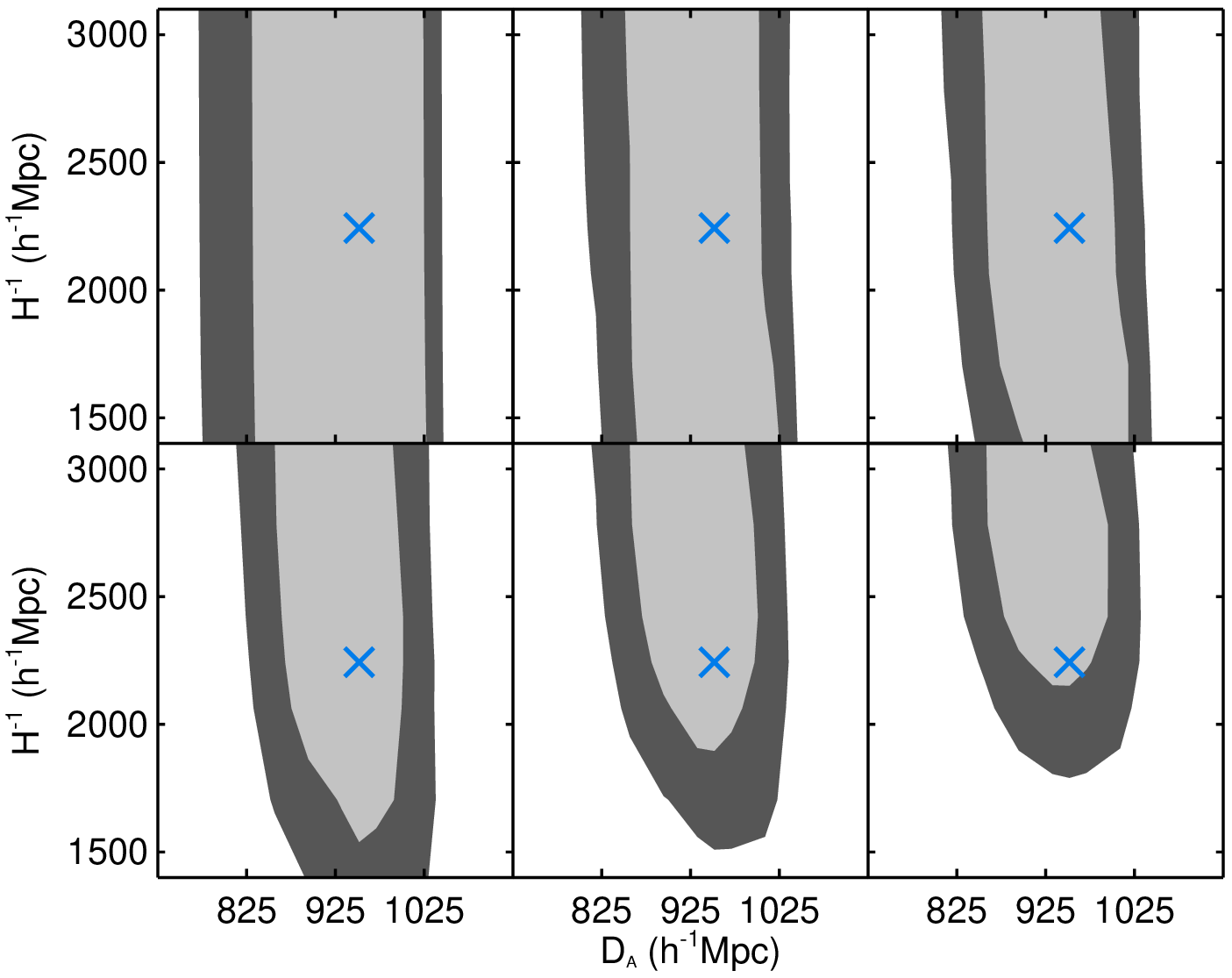} 
	\caption{\textit{Left panel:} Constraints at the 68 and 95 per cent confidence limit on the
			  parameters $D_{A}$ and $H^{-1}$ obtained from the spectroscopic sample. 
			  The $\chi^{2}_{m}$ values used are from $\bar \mu=0.08$ (top left), 
			  $\bar \mu=0.08+0.25$ (top middle), $\bar \mu=0.08+0.25+0.42$
			  (top right), $\bar \mu=0.08+0.25+0.42+0.58$ (bottom left),
			  $\bar \mu=0.08+0.25+0.42+0.58+0.75$ (bottom middle) and 
			  $\bar \mu=0.08+0.25+0.42+0.58+0.75+0.92$ (bottom right). 
			  The blue cross marks the fiducial value of $D_{A}$ and $H^{-1}$.		 
			  Units are in $\mathrm{h}^{-1}\mathrm{Mpc}$. 
			  \textit{Right panel:} The constraints obtained from the \sigmaone photometric 
			  sample with the same methodology of adding up the $\chi^{2}_{m}$ values. }
	\label{fig:DA_vs_Hinv_zspec}   
\end{figure*}

There are challenges in computing the theoretical prediction of galaxy clustering in redshift 
space. Although cosmic distances are estimated using the BAO at linear regimes, the small 
peak structure tends to be smeared out by non--linear physics which needs to be computed. 
In addition, the infinite higher order polynomials are generated due to the density and 
velocity correlations. Those perturbative corrections in a more elaborate description are 
included to make precise prediction of the BAO structure~\citep{Taruya:2010mx}. Finally, 
those perturbative effects and non--linear smearing effects on the clustering are not 
separately modelled. Taking account of this fact, \cite{Taruya:2010mx} proposed an improved 
model of the redshift-space power spectrum, in which the coupling between the density and 
velocity fields associated with the Kaiser and the FoG effects is perturbatively incorporated 
into the power spectrum expression. The resultant includes nonlinear corrections consisting of 
higher-order polynomials~\citep{Taruya:2010mx}:
\ba
\tilde{P}(k,\mu) &=& \big\{P_{\delta\delta}(k) + 2\mu^2 P_{\delta\Theta}(k) + \mu^4 
P_{\Theta\Theta}(k) \nn\\
&+& A(k,\mu) + B(k,\mu)\big\}G^{\rm FoG}
\label{eq:TNS10}
\ea
Here the $A(k,\mu)$ and $B(k,\mu)$ terms are the nonlinear corrections, and are expanded as 
power series of $\mu$. Those spectra are computed using the fiducial cosmological parameters. 
The FoG effect $G^{\rm FoG}$ is given by the simple Gaussian function which is written as,
\ba
G^{\rm FoG} \equiv  \exp{\left[-(k\mu\sigma_p)^2\right]}
\ea
where $\sigma_p$ denotes one dimensional velocity dispersion. Thus the theoretical correlation 
function $\xi_{th}$ is parameterised by ($D_{A},H^{-1},G_{b},G_{\Theta},\sigma_{p}$) wherein 
$G_{b}$ and $G_{\Theta}$ are the normalised density and coherent motion growth functions. The 
BAO feature is weakly dependent on the growth functions and $\sigma_{p}$. When working with 
spectroscopic redshift samples for which the error on the redshift is negligible, we can 
marginalise over the above set of 5 parameters and the corresponding $\xi_{th}(s,\mu)$ can be 
used as the fit to the observed \xismu. But when working with photo-z samples, the effect of 
the photo-z error on the correlation function is incoherent. Thus, the extra parameter needed 
for the theoretical template to model $\xi_{th}(s,\mu)$ as a function of the photo-z error is 
not well understood. So, we use Eq. \ref{eqn:empirical_fit} instead to fit our observed \xismu.
This functional form only assumes a power-law at small scales and a Gaussian function to fit 
the BAO peak at large scales and seems to model \xismu quite well as we can see from Fig.
\ref{fig:s_vs_xismu_zspec_vs_zphot_all}. The effect of the photo-z error on $\xi_{th}(s,\mu)$ 
and its marginalisation is kept for future work.

In this verification work, when we fit the cosmic distances, we vary the tangential and 
radial distance measures from the fiducial values of 
$D_{A}=951.80 \,\mathrm{h}^{-1}\mathrm{Mpc}$ and 
$H^{-1}=2243.04 \,\mathrm{h}^{-1}\mathrm{Mpc}$. The best fit $\sigma_p$ for this simulation 
is found to be $\sigma_{p}=4.2\,\mathrm{h}^{-1}\mathrm{Mpc}$. 
Note that we apply the TNS model for computing the theoretical BAO peaks to fit the measured 
data. The theoretical BAO peaks at the fiducial cosmology can be transformed into a new 
cosmology according to the simple coordinate projections, which are presented as solid curves 
in Fig.~\ref{fig:DA_Hinv_var}. But the location shift of BAO peak with varying $D_A$ and 
$H^{-1}$ is not completely consistent with coordinate transformation. 
The theoretical BAO peaks computed using the TNS model are represented by dotted points. 
The difference is exceeding the 
dectectability limit by about 5\%, and thus the TNS model is adopted to determine the 
theoretical BAO points.

\begin{figure*}
	\includegraphics[width=3.4in]{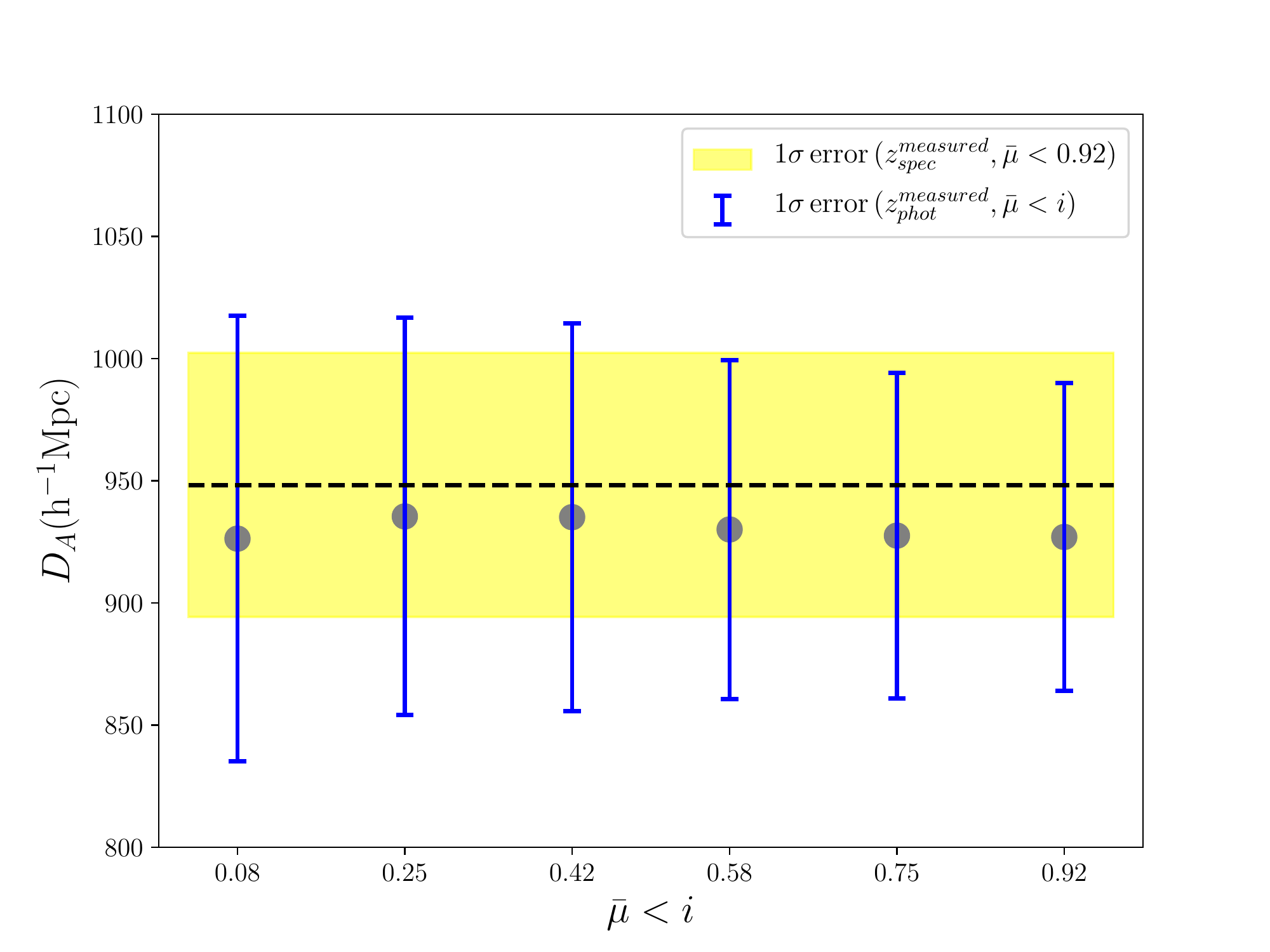}  
	\includegraphics[width=3.4in]{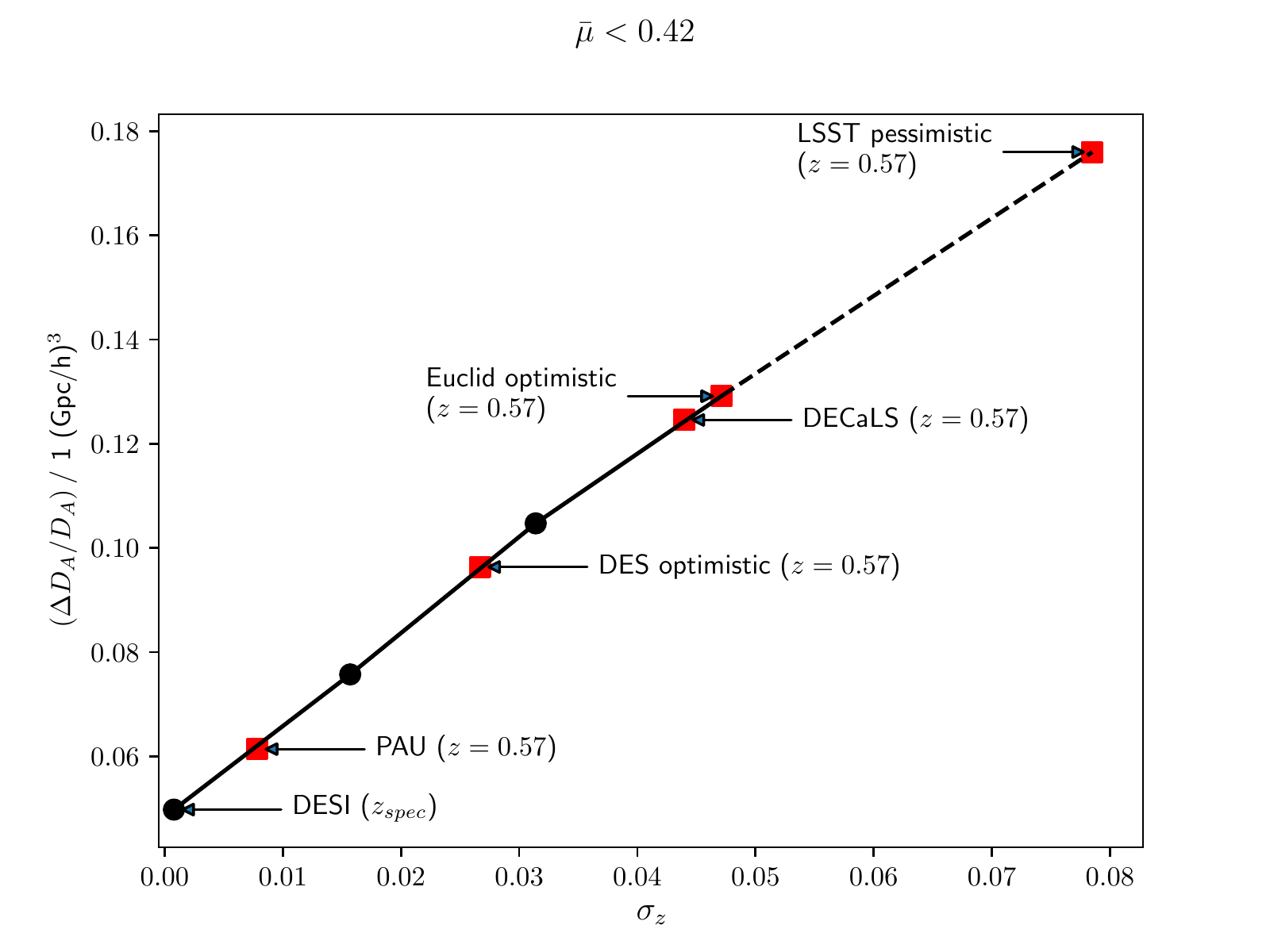} 
	\caption{\textit{Left panel:} The measured $D_{A}$ and the corresponding $1\sigma$ errors
	         calculated using $\chi^{2}_{m}(\bar \mu < i)$, where $i$
		     goes from 0.08 to 0.92, for the $\sigma_{0} = 0.01$ photometric sample (plotted
		     using the blue dots). The dashed black line denotes the measured $D_{A}$
		     using $\chi^{2}_{m}(\bar \mu < 0.92)$ for the spectroscopic sample 
		     along with the yellow highlighted region denoting the $1\sigma$ errors.
			 \textit{Right panel:} $\sigma_{z}$ vs $\Delta D_{A}/D_{A}$ per 1 (Gpc$/h)^{3}$ 
			 for $\bar \mu <0.42$. All the values of $\sigma_{z}$ correspond to the 
			 median redshift $\bar z = 0.57$. The three black dots are values obtained for 
			 the spectroscopic redshift, $\sigma_{0} = 0.01$ and 0.02 respectively. The 
			 red squares give us values of $\Delta D_{A}/D_{A}$ for different 
			 surveys based on their expected $\sigma_{0}$. The dotted black line is a crude 
			 approximation out to $\sigma_{0} = 0.05$ corresponding to the LSST 
			 pessimistic case.}
	\label{fig:mu_vs_DA_zspec_vs_zphot}   
\end{figure*}

\subsection{Measured cosmic distances}
\label{sec:DA_Hinv_constraints}

With the given fiducial cosmology, both the tangential and radial 
BAO peak locations can be computed by varying $D_{A}(z)$ and $H^{-1}(z)$ using the theoretical 
templates introduced in the previous subsection. The cosmic distances estimated from the 
measured BAO peak locations at all 6 $\mu$ bins are shown in Fig.\ref{fig:DA_vs_Hinv_zspec}.
Different combinations of $\mu_i$ bins in which all bins of $\mu_i$ with $i<i_{\rm max}$ are 
cumulatively summed and the $i_{\rm max}$ runs from 1 at the top left to 6 at the bottom 
right.

In principle, because the BAO ring spans both the transverse and radial cosmological 
coordinates, it can be exploited to probe the distance measures of $D_{A}(z)$ and $H^{-1}(z)$ 
separately. If there is minimal photo-z uncertainty, then both cosmic distances are precisely 
measured as presented in the left panel of Fig.\ref{fig:DA_vs_Hinv_zspec}. The uncertainty 
on the redshift determination prevents us from accessing the radial cosmic distance, and thus 
$H^{-1}(z)$ is poorly determined as presented in the right panel. However, note that the 
transverse distance is measured precisely regardless of the systematic uncertainty along the 
radial direction. Although $D_{A}(z)$ is measured after full marginalisation over $H^{-1}(z)$, 
the precision loss in the $D_{A}(z)$ measurement is negligible, which can be compared between 
the left and right panels of Fig.\ref{fig:DA_vs_Hinv_zspec}. 
The constraints obtained on the fiducial and the measured $D_{A}(\bar{z}=0.57)$ from the 
photometric sample when $i<(i_{\rm max}=6)$ are 
$D_{A}(\bar{z}=0.57)=951.80^{+37.40}_{-87.80}\, \mathrm{h}^{-1}\mathrm{Mpc}$ and 
$D_{A}(\bar{z}=0.57)=927.00^{+63.00}_{-63.01}\, \mathrm{h}^{-1}\mathrm{Mpc}$ respectively. 
The constraints on $D_{A}(z)$ and $H^{-1}(z)$ improve with increasing $i_{\rm max}$. 
The error 
on $D_{A}(z)$ decreases from 9.5\% (for $i_{\rm max}=1$) to 5.7\% (for $i_{\rm max}=6)$ 
for the spectroscopic sample and a similar trend is observed for the \sigmaone 
photometric sample, with the error on $D_{A}(z)$ decreasing from 9.8\% to 6.5\% as shown 
in the left panel of Fig.\ref{fig:mu_vs_DA_zspec_vs_zphot}. 

The constraints obtained on $D_{A}(z)$ not only depend on $\sigma_{0}$ and $z$, 
but also on
the volume covered by the survey. In the right panel of Fig.\ref{fig:mu_vs_DA_zspec_vs_zphot}
we plot $\sigma_{z}$ vs $\Delta D_{A}/D_{A}$ per 1 (Gpc$/h)^{3}$ at $\bar z=0.57$ and
$\bar \mu < 0.42$
for the spectroscopic and the two ($\sigma_{0} = 0.01, 0.02$) photometric cases which 
are denoted by the black dots. We also plot the $\Delta D_{A}/D_{A}$ for different surveys
(red squares) based on their expected $\sigma_{0}$, starting with 0.005 for the PAUS survey 
\citep{Benitez_2009_PAU,Marti_PAU}, 0.017 for DES Y1 high luminosity `redmagic'
\citep{Rozo_2016_DES} sample, 0.028 (Zhou. et al 2019, in preparation) for DECaLS 
\citep{Decals} LRGs and 0.030 for \textit{Euclid} \citep{euclid_red_book}. The black dotted 
line is a crude approximation out to $\sigma_{0} = 0.05$ which corresponds to the LSST
pessimistic case \citep{LSST_1,LSST_2}.

\section{Discussion and conclusions}\label{sec:conclusion}

We study the statistical methodology to extract the cosmic distance information from 
photometric galaxy samples, using the simulated photometric galaxy distribution based on the 
DR12 CMASS map. The measured monopole moment of the two-point correlation function is so 
significantly contaminated by the uncertainty in redshift determination that the BAO feature 
is smeared out completely. 
The common practice to extract the BAO peak is to exploit the incomplete angular 
averaged correlation which is known as the projected correlation function. When the most 
contaminated correlation configuration along the LOS is removed, the BAO peak starts becoming 
visible.

In this manuscript, the wedge is binned into 6 pieces using equal $\mu$ spacing from $\mu=0$ 
to 1 and we analyse each wedge component one by one and present the level of 
contamination due to the $z$ uncertainty, in comparison to the spectroscopic map. 
We find that the first two wedge correlations are least contaminated by the $z$ uncertainty. 
The noticeable contamination 
is observed from the third $\mu$ bin with the BAO peak still visible. The transverse cosmic 
distance is probed with a reasonably good precision as presented in 
Fig.\ref{fig:mu_vs_DA_zspec_vs_zphot}. Those wedges are coherently summed using the full 
covariance matrix, and the cumulative constraint on $D_A$ is presented to show the information 
is nearly saturated at the first several bins. For the radial component of the 
cosmic distance, we 
are not able to extract it from the photometric sample, as most radial information is 
contained 
at wedge bins higher than $\mu_i$ with $i\ga 4$, which is contaminated by the $z$ uncertainty. 
However, the measured transverse cosmic distance is immune from this uncertainty. The reported 
values of $D_A$ in Fig.\ref{fig:mu_vs_DA_zspec_vs_zphot} 
is computed after full marginalisation 
of $H^{-1}$. The measured $D_A$ is not biased by this marginalisation.

Multiband imaging surveys rely on photometric redshift for radial information. 
The Dark Energy Survey (DES) \citep{DES} and future surveys such as 
LSST \citep{LSST_1,LSST_2} and \textit{Euclid} \citep{euclid_red_book} will provide 
state-of-the-art photometric redshifts over an unprecedented range of redshift scales. 
The precision that is expected from the photometric redshifts is typically 3\% 
\citep{Rozo_2016_DES}. Some of the recent works that have used photometric redshift catalogues 
have focused on measuring the angular correlation function $w(\theta)$ to get cosmic 
distance measurements \citep{Sanchez_2011,Seo_2012,Carnero_2012} using several narrow 
redshift slices. However, they clearly do not use the full information available, as 
radial binning blends data beyond what is induced by the photometric redshift error. 
Another important aspect that is often ignored when calculating $w(\theta)$ is cross 
correlation between the different redshift bins used. Using many redshift slices also 
complicates the computation of the covariance matrix, with the computing time increasing with 
the number of bins in $\theta$ and number of redshift slices used. It has also been shown 
recently by \cite{Ross_2017} that the statistics obtained using \xismu are about 6\% more 
accurate compared to $w(\theta)$. Thus, using \xismu not only adds more information 
compared to 
$w(\theta)$, but also overcomes the above disadvantages.

The full photometric footprints for DESI were released ahead of the spectroscopic follow up. 
We have shown here that the transverse component of cosmic distance can be pre--measured using 
the photometric galaxy map even without the need of a spectroscopic follow up in the future. 
In addition, the photometric survey leads us to deeper redshifts which will not be probed by 
the spectroscopic survey. For instance, the spectroscopic LRG sample ends at redshift 
$z\sim 0.8$, but the photometric sample reaches up to $z\ga 1.0$. 
We verify in this manuscript that we are able to probe the transverse distance at a higher 
redshift using the photometric map. In our next project, this verified method will be applied 
to measure cosmic distances for the DECaLS \citep{Decals} 
DR7 photometric LRG galaxy map (Zhou. et al 2019, in preparation) at redshift ranges of 
$0.6<z_{phot}<0.8$ and $0.8<z_{phot}<1.0$. 
The spectroscopic follow up survey fully covers LRG galaxies in 
the range of $0.6<z<0.8$, and partially for the $0.8<z<1.0$ case. If the cosmic distance at 
$0.8<z<1.0$ is successfully measured, 
then we will be able to probe cosmological information that is not fully covered by the 
spectroscopic survey.


\section*{Acknowledgements}
 
Data analysis was performed using the high performance computing cluster \textit{POLARIS} 
at the Korea Astronomy and Space Science Institute. 
This research made use of TOPCAT and STIL: Starlink Table/VOTable Processing Software 
developed by \citet{topcat} 
and also the Code for Anisotropies in the Microwave Background (CAMB) \citep{CAMB_1,CAMB_2}.
Srivatsan Sridhar would also like to thank Sridhar Krishnan, Revathy Sridhar and 
Madhumitha Srivatsan for their support and encouragement during this work.




\bibliographystyle{mnras}
\bibliography{YS_09_Jan} 

\begin{thebibliography}{}
\makeatletter
\relax
\def\mn@urlcharsother{\let\do\@makeother \do\$\do\&\do\#\do\^\do\_\do\%\do\~}
\def\mn@doi{\begingroup\mn@urlcharsother \@ifnextchar [ {\mn@doi@}
  {\mn@doi@[]}}
\def\mn@doi@[#1]#2{\def\@tempa{#1}\ifx\@tempa\@empty \href
  {http://dx.doi.org/#2} {doi:#2}\else \href {http://dx.doi.org/#2} {#1}\fi
  \endgroup}
\def\mn@eprint#1#2{\mn@eprint@#1:#2::\@nil}
\def\mn@eprint@arXiv#1{\href {http://arxiv.org/abs/#1} {{\tt arXiv:#1}}}
\def\mn@eprint@dblp#1{\href {http://dblp.uni-trier.de/rec/bibtex/#1.xml}
  {dblp:#1}}
\def\mn@eprint@#1:#2:#3:#4\@nil{\def\@tempa {#1}\def\@tempb {#2}\def\@tempc
  {#3}\ifx \@tempc \@empty \let \@tempc \@tempb \let \@tempb \@tempa \fi \ifx
  \@tempb \@empty \def\@tempb {arXiv}\fi \@ifundefined
  {mn@eprint@\@tempb}{\@tempb:\@tempc}{\expandafter \expandafter \csname
  mn@eprint@\@tempb\endcsname \expandafter{\@tempc}}}

\bibitem[\protect\citeauthoryear{{Alam} et~al.,}{{Alam}
  et~al.}{2017}]{Alam_2017}
{Alam} S.,  et~al., 2017, \mn@doi [\mnras] {10.1093/mnras/stx721}, \href
  {http://adsabs.harvard.edu/abs/2017MNRAS.470.2617A} {470, 2617}

\bibitem[\protect\citeauthoryear{{Ascaso}, {Mei}  \& {Ben{\'{\i}}tez}}{{Ascaso}
  et~al.}{2015}]{Ascaso}
{Ascaso} B.,  {Mei} S.,   {Ben{\'{\i}}tez} N.,  2015, \mn@doi [\mnras]
  {10.1093/mnras/stv1597}, \href
  {http://adsabs.harvard.edu/abs/2015MNRAS.453.2515A} {453, 2515}

\bibitem[\protect\citeauthoryear{{Benedict} et~al.,}{{Benedict}
  et~al.}{1999}]{Parallax}
{Benedict} G.~F.,  et~al., 1999, \mn@doi [\aj] {10.1086/300975}, \href
  {http://adsabs.harvard.edu/abs/1999AJ....118.1086B} {118, 1086}

\bibitem[\protect\citeauthoryear{{Ben{\'{\i}}tez} et~al.,}{{Ben{\'{\i}}tez}
  et~al.}{2009}]{Benitez_2009_PAU}
{Ben{\'{\i}}tez} N.,  et~al., 2009, \mn@doi [\apj]
  {10.1088/0004-637X/691/1/241}, \href
  {https://ui.adsabs.harvard.edu/abs/2009ApJ...691..241B} {691, 241}

\bibitem[\protect\citeauthoryear{{Benitez} et~al.,}{{Benitez}
  et~al.}{2014}]{JPAS}
{Benitez} N.,  et~al., 2014, arXiv e-prints, \href
  {https://ui.adsabs.harvard.edu/abs/2014arXiv1403.5237B} {p. arXiv:1403.5237}

\bibitem[\protect\citeauthoryear{{Blake} \& {Glazebrook}}{{Blake} \&
  {Glazebrook}}{2003}]{Blake_2003}
{Blake} C.,  {Glazebrook} K.,  2003, \mn@doi [\apj] {10.1086/376983}, \href
  {http://adsabs.harvard.edu/abs/2003ApJ...594..665B} {594, 665}

\bibitem[\protect\citeauthoryear{{Carnero}, {S{\'a}nchez}, {Crocce},
  {Cabr{\'e}}  \& {Gazta{\~n}aga}}{{Carnero} et~al.}{2012}]{Carnero_2012}
{Carnero} A.,  {S{\'a}nchez} E.,  {Crocce} M.,  {Cabr{\'e}} A.,
  {Gazta{\~n}aga} E.,  2012, \mn@doi [\mnras]
  {10.1111/j.1365-2966.2011.19832.x}, \href
  {http://adsabs.harvard.edu/abs/2012MNRAS.419.1689C} {419, 1689}

\bibitem[\protect\citeauthoryear{{DESI Collaboration} et~al.,}{{DESI
  Collaboration} et~al.}{2016}]{DESI_1}
{DESI Collaboration} et~al., 2016, arXiv e-prints, \href
  {https://ui.adsabs.harvard.edu/abs/2016arXiv161100036D} {p. arXiv:1611.00036}

\bibitem[\protect\citeauthoryear{{Davis} \& {Peebles}}{{Davis} \&
  {Peebles}}{1983}]{davis_peebles_1983}
{Davis} M.,  {Peebles} P.~J.~E.,  1983, \mn@doi [\apj] {10.1086/160884}, \href
  {http://adsabs.harvard.edu/abs/1983ApJ...267..465D} {267, 465}

\bibitem[\protect\citeauthoryear{{Dey} et~al.,}{{Dey} et~al.}{2018}]{Decals}
{Dey} A.,  et~al., 2018, preprint, \href
  {http://adsabs.harvard.edu/abs/2018arXiv180408657D} {} (\mn@eprint {arXiv}
  {1804.08657})

\bibitem[\protect\citeauthoryear{{Eisenstein}, {Hu}  \& {Tegmark}}{{Eisenstein}
  et~al.}{1998}]{Eisenstein_1998}
{Eisenstein} D.~J.,  {Hu} W.,   {Tegmark} M.,  1998, \mn@doi [\apjl]
  {10.1086/311582}, \href {http://adsabs.harvard.edu/abs/1998ApJ...504L..57E}
  {504, L57}

\bibitem[\protect\citeauthoryear{{Eisenstein} et~al.,}{{Eisenstein}
  et~al.}{2005}]{Eisenstein_2005}
{Eisenstein} D.~J.,  et~al., 2005, \mn@doi [\apj] {10.1086/466512}, \href
  {http://adsabs.harvard.edu/abs/2005ApJ...633..560E} {633, 560}

\bibitem[\protect\citeauthoryear{{Estrada}, {Sefusatti}  \&
  {Frieman}}{{Estrada} et~al.}{2009}]{Estrada_2009}
{Estrada} J.,  {Sefusatti} E.,   {Frieman} J.~A.,  2009, \mn@doi [\apj]
  {10.1088/0004-637X/692/1/265}, \href
  {http://adsabs.harvard.edu/abs/2009ApJ...692..265E} {692, 265}

\bibitem[\protect\citeauthoryear{{Farrow} et~al.,}{{Farrow}
  et~al.}{2015}]{Farrow}
{Farrow} D.~J.,  et~al., 2015, \mn@doi [\mnras] {10.1093/mnras/stv2075}, \href
  {http://adsabs.harvard.edu/abs/2015MNRAS.454.2120F} {454, 2120}

\bibitem[\protect\citeauthoryear{{Fernie}}{{Fernie}}{1969}]{Standard_candle}
{Fernie} J.~D.,  1969, \mn@doi [\pasp] {10.1086/128847}, \href
  {http://adsabs.harvard.edu/abs/1969PASP...81..707F} {81, 707}

\bibitem[\protect\citeauthoryear{{Hong}, {Han}, {Wen}, {Sun}  \& {Zhan}}{{Hong}
  et~al.}{2012}]{Hong_2012}
{Hong} T.,  {Han} J.~L.,  {Wen} Z.~L.,  {Sun} L.,   {Zhan} H.,  2012, \mn@doi
  [\apj] {10.1088/0004-637X/749/1/81}, \href
  {http://adsabs.harvard.edu/abs/2012ApJ...749...81H} {749, 81}

\bibitem[\protect\citeauthoryear{Howlett, Lewis, Hall  \& Challinor}{Howlett
  et~al.}{2012}]{CAMB_2}
Howlett C.,  Lewis A.,  Hall A.,   Challinor A.,  2012, \mn@doi [JCAP]
  {10.1088/1475-7516/2012/04/027}, 1204, 027

\bibitem[\protect\citeauthoryear{{Ivezic} et~al.,}{{Ivezic}
  et~al.}{2008}]{LSST_1}
{Ivezic} Z.,  et~al., 2008, preprint, \href
  {http://adsabs.harvard.edu/abs/2008arXiv0805.2366I} {} (\mn@eprint {arXiv}
  {0805.2366})

\bibitem[\protect\citeauthoryear{{Kazin} et~al.,}{{Kazin}
  et~al.}{2013}]{Kazin_2013}
{Kazin} E.~A.,  et~al., 2013, \mn@doi [\mnras] {10.1093/mnras/stt1261}, \href
  {http://adsabs.harvard.edu/abs/2013MNRAS.435...64K} {435, 64}

\bibitem[\protect\citeauthoryear{{Kerscher}, {Szapudi}  \& {Szalay}}{{Kerscher}
  et~al.}{2000}]{kerscher}
{Kerscher} M.,  {Szapudi} I.,   {Szalay} A.~S.,  2000, \mn@doi [\apj]
  {10.1086/312702}, \href {http://adsabs.harvard.edu/abs/2000ApJ...535..13K}
  {535, 13}

\bibitem[\protect\citeauthoryear{{LSST Science Collaboration} et~al.,}{{LSST
  Science Collaboration} et~al.}{2009}]{LSST_2}
{LSST Science Collaboration} et~al., 2009, preprint, \href
  {http://adsabs.harvard.edu/abs/2009arXiv0912.0201L} {} (\mn@eprint {arXiv}
  {0912.0201})

\bibitem[\protect\citeauthoryear{{Laureijs} et~al.,}{{Laureijs}
  et~al.}{2011}]{euclid_red_book}
{Laureijs} R.,  et~al., 2011, preprint, \href
  {http://adsabs.harvard.edu/abs/2011arXiv1110.3193L} {} (\mn@eprint {arXiv}
  {1110.3193})

\bibitem[\protect\citeauthoryear{Lewis, Challinor  \& Lasenby}{Lewis
  et~al.}{2000}]{CAMB_1}
Lewis A.,  Challinor A.,   Lasenby A.,  2000, \mn@doi [Astrophys. J.]
  {10.1086/309179}, 538, 473

\bibitem[\protect\citeauthoryear{{Manera} et~al.,}{{Manera}
  et~al.}{2013}]{Manera}
{Manera} M.,  et~al., 2013, \mn@doi [\mnras] {10.1093/mnras/sts084}, \href
  {http://adsabs.harvard.edu/abs/2013MNRAS.428.1036M} {428, 1036}

\bibitem[\protect\citeauthoryear{{Mart{\'{\i}}}, {Miquel}, {Castander},
  {Gazta{\~n}aga}, {Eriksen}  \& {S{\'a}nchez}}{{Mart{\'{\i}}}
  et~al.}{2014}]{Marti_PAU}
{Mart{\'{\i}}} P.,  {Miquel} R.,  {Castander} F.~J.,  {Gazta{\~n}aga} E.,
  {Eriksen} M.,   {S{\'a}nchez} C.,  2014, \mn@doi [\mnras]
  {10.1093/mnras/stu801}, \href
  {https://ui.adsabs.harvard.edu/abs/2014MNRAS.442...92M} {442, 92}

\bibitem[\protect\citeauthoryear{{Padilla} et~al.,}{{Padilla}
  et~al.}{2019}]{PAU}
{Padilla} C.,  et~al., 2019, arXiv e-prints, \href
  {https://ui.adsabs.harvard.edu/abs/2019arXiv190203623P} {p. arXiv:1902.03623}

\bibitem[\protect\citeauthoryear{{Padmanabhan}, {Xu}, {Eisenstein}, {Scalzo},
  {Cuesta}, {Mehta}  \& {Kazin}}{{Padmanabhan} et~al.}{2012}]{Padmanabhan_2012}
{Padmanabhan} N.,  {Xu} X.,  {Eisenstein} D.~J.,  {Scalzo} R.,  {Cuesta} A.~J.,
   {Mehta} K.~T.,   {Kazin} E.,  2012, \mn@doi [\mnras]
  {10.1111/j.1365-2966.2012.21888.x}, \href
  {http://adsabs.harvard.edu/abs/2012MNRAS.427.2132P} {427, 2132}

\bibitem[\protect\citeauthoryear{{Peebles} \& {Yu}}{{Peebles} \&
  {Yu}}{1970}]{Peebles_Yu}
{Peebles} P.~J.~E.,  {Yu} J.~T.,  1970, \mn@doi [\apj] {10.1086/150713}, \href
  {http://adsabs.harvard.edu/abs/1970ApJ...162..815P} {162, 815}

\bibitem[\protect\citeauthoryear{{Perlmutter} et~al.,}{{Perlmutter}
  et~al.}{1999}]{Perlmutter_1999}
{Perlmutter} S.,  et~al., 1999, \mn@doi [\apj] {10.1086/307221}, \href
  {http://adsabs.harvard.edu/abs/1999ApJ...517..565P} {517, 565}

\bibitem[\protect\citeauthoryear{{Riess} et~al.,}{{Riess}
  et~al.}{1998}]{Riess_1998}
{Riess} A.~G.,  et~al., 1998, \mn@doi [\aj] {10.1086/300499}, \href
  {http://adsabs.harvard.edu/abs/1998AJ....116.1009R} {116, 1009}

\bibitem[\protect\citeauthoryear{{Ross} et~al.,}{{Ross}
  et~al.}{2012}]{Ross_2012}
{Ross} A.~J.,  et~al., 2012, \mn@doi [\mnras]
  {10.1111/j.1365-2966.2012.21235.x}, \href
  {http://adsabs.harvard.edu/abs/2012MNRAS.424..564R} {424, 564}

\bibitem[\protect\citeauthoryear{{Ross} et~al.,}{{Ross}
  et~al.}{2017}]{Ross_2017}
{Ross} A.~J.,  et~al., 2017, \mn@doi [\mnras] {10.1093/mnras/stx2120}, \href
  {https://ui.adsabs.harvard.edu/#abs/2017MNRAS.472.4456R} {472, 4456}

\bibitem[\protect\citeauthoryear{{Rozo} et~al.,}{{Rozo}
  et~al.}{2016}]{Rozo_2016_DES}
{Rozo} E.,  et~al., 2016, \mn@doi [\mnras] {10.1093/mnras/stw1281}, \href
  {https://ui.adsabs.harvard.edu/abs/2016MNRAS.461.1431R} {461, 1431}

\bibitem[\protect\citeauthoryear{{Sabiu}}{{Sabiu}}{2018}]{KSTAT}
{Sabiu} C.,  2018, {KSTAT: KD-tree Statistics Package}, Astrophysics Source
  Code Library (\mn@eprint {ascl} {1804.026})

\bibitem[\protect\citeauthoryear{{Sabiu} \& {Song}}{{Sabiu} \&
  {Song}}{2016}]{Cris_2016}
{Sabiu} C.~G.,  {Song} Y.-S.,  2016, preprint, \href
  {http://adsabs.harvard.edu/abs/2016arXiv160302389S} {} (\mn@eprint {arXiv}
  {1603.02389})

\bibitem[\protect\citeauthoryear{{S{\'a}nchez} et~al.,}{{S{\'a}nchez}
  et~al.}{2011}]{Sanchez_2011}
{S{\'a}nchez} E.,  et~al., 2011, \mn@doi [\mnras]
  {10.1111/j.1365-2966.2010.17679.x}, \href
  {http://adsabs.harvard.edu/abs/2011MNRAS.411..277S} {411, 277}

\bibitem[\protect\citeauthoryear{{S{\'a}nchez} et~al.,}{{S{\'a}nchez}
  et~al.}{2012}]{Sanchez_empirical}
{S{\'a}nchez} A.~G.,  et~al., 2012, \mn@doi [\mnras]
  {10.1111/j.1365-2966.2012.21502.x}, \href
  {http://adsabs.harvard.edu/abs/2012MNRAS.425..415S} {425, 415}

\bibitem[\protect\citeauthoryear{{S{\'a}nchez} et~al.,}{{S{\'a}nchez}
  et~al.}{2014a}]{Sanchez_2014}
{S{\'a}nchez} A.~G.,  et~al., 2014a, \mn@doi [\mnras] {10.1093/mnras/stu342},
  \href {http://adsabs.harvard.edu/abs/2014MNRAS.440.2692S} {440, 2692}

\bibitem[\protect\citeauthoryear{{S{\'a}nchez} et~al.,}{{S{\'a}nchez}
  et~al.}{2014b}]{Sanchez_DES}
{S{\'a}nchez} C.,  et~al., 2014b, \mn@doi [\mnras] {10.1093/mnras/stu1836},
  \href {http://adsabs.harvard.edu/abs/2014MNRAS.445.1482S} {445, 1482}

\bibitem[\protect\citeauthoryear{{S{\'a}nchez} et~al.,}{{S{\'a}nchez}
  et~al.}{2017}]{Sanchez_2017}
{S{\'a}nchez} A.~G.,  et~al., 2017, \mn@doi [\mnras] {10.1093/mnras/stw2443},
  \href {http://adsabs.harvard.edu/abs/2017MNRAS.464.1640S} {464, 1640}

\bibitem[\protect\citeauthoryear{{Seo} et~al.,}{{Seo} et~al.}{2012}]{Seo_2012}
{Seo} H.-J.,  et~al., 2012, \mn@doi [\apj] {10.1088/0004-637X/761/1/13}, \href
  {http://adsabs.harvard.edu/abs/2012ApJ...761...13S} {761, 13}

\bibitem[\protect\citeauthoryear{{Sridhar}, {Maurogordato}, {Benoist}, {Cappi}
  \& {Marulli}}{{Sridhar} et~al.}{2017}]{Srivatsan}
{Sridhar} S.,  {Maurogordato} S.,  {Benoist} C.,  {Cappi} A.,   {Marulli} F.,
  2017, \mn@doi [\aap] {10.1051/0004-6361/201629369}, \href
  {http://adsabs.harvard.edu/abs/2017A%26A...600A..32S} {600, A32}

\bibitem[\protect\citeauthoryear{{Taruya}, {Nishimichi}  \& {Saito}}{{Taruya}
  et~al.}{2010}]{Taruya:2010mx}
{Taruya} A.,  {Nishimichi} T.,   {Saito} S.,  2010, \mn@doi [\prd]
  {10.1103/PhysRevD.82.063522}, \href
  {http://adsabs.harvard.edu/abs/2010PhRvD..82f3522T} {82, 063522}

\bibitem[\protect\citeauthoryear{{Taruya}, {Bernardeau}, {Nishimichi}  \&
  {Codis}}{{Taruya} et~al.}{2012}]{Taruya:2012ut}
{Taruya} A.,  {Bernardeau} F.,  {Nishimichi} T.,   {Codis} S.,  2012, \mn@doi
  [\prd] {10.1103/PhysRevD.86.103528}, \href
  {http://adsabs.harvard.edu/abs/2012PhRvD..86j3528T} {86, 103528}

\bibitem[\protect\citeauthoryear{{Taylor}}{{Taylor}}{2005}]{topcat}
{Taylor} M.~B.,  2005, in {Shopbell} P.,  {Britton} M.,   {Ebert} R.,  eds,
  Astronomical Society of the Pacific Conference Series Vol. 347, Astronomical
  Data Analysis Software and Systems XIV. p.~29

\bibitem[\protect\citeauthoryear{{The Dark Energy Survey Collaboration}}{{The
  Dark Energy Survey Collaboration}}{2005}]{DES}
{The Dark Energy Survey Collaboration} 2005, ArXiv Astrophysics e-prints, \href
  {http://adsabs.harvard.edu/abs/2005astro.ph.10346T} {}

\bibitem[\protect\citeauthoryear{{Totsuji} \& {Kihara}}{{Totsuji} \&
  {Kihara}}{1969}]{totsuji_1969}
{Totsuji} H.,  {Kihara} T.,  1969, \pasj, \href
  {http://adsabs.harvard.edu/abs/1969PASJ...21..221T} {21, 221}

\bibitem[\protect\citeauthoryear{{Veropalumbo}, {Marulli}, {Moscardini},
  {Moresco}  \& {Cimatti}}{{Veropalumbo} et~al.}{2014}]{Veropalumbo_2014}
{Veropalumbo} A.,  {Marulli} F.,  {Moscardini} L.,  {Moresco} M.,   {Cimatti}
  A.,  2014, \mn@doi [\mnras] {10.1093/mnras/stu1050}, \href
  {http://adsabs.harvard.edu/abs/2014MNRAS.442.3275V} {442, 3275}

\bibitem[\protect\citeauthoryear{{Veropalumbo}, {Marulli}, {Moscardini},
  {Moresco}  \& {Cimatti}}{{Veropalumbo} et~al.}{2016}]{Veropalumbo_2016}
{Veropalumbo} A.,  {Marulli} F.,  {Moscardini} L.,  {Moresco} M.,   {Cimatti}
  A.,  2016, \mn@doi [\mnras] {10.1093/mnras/stw306}, \href
  {http://adsabs.harvard.edu/abs/2016MNRAS.458.1909V} {458, 1909}

\makeatother
\end{thebibliography}




\appendix
\bsp	
\label{lastpage}
\end{document}